\documentclass[apj]{emulateapj}
\pdfoutput=1
\submitted{ApJ, accepted 17/01/2014}

\usepackage{amsmath}
\usepackage{amsfonts}
\usepackage{amsthm}
\usepackage{amssymb}
\usepackage{graphicx}
\usepackage{mathrsfs}
\usepackage{latexsym}
\usepackage{graphics}
\usepackage{cases}
\usepackage{hyperref}
\usepackage{mathtools}
\usepackage{dsfont}
\usepackage[percent]{overpic}
\usepackage{comment} 
\usepackage{enumerate}

\newcommand{\be}{\begin{equation}}
\newcommand{\ee}{\end{equation}}
\newcommand{\bea}{\begin{eqnarray}}
\newcommand{\eea}{\end{eqnarray}}
\newcommand{\bes}{\begin{equation}\begin{split}}
\newcommand{\ees}{{\end{split}\end{equation}}}
\newcommand{\eq}[1]{Equation~(\ref{eq_#1})}
\newcommand{\fig}[1]{Figure~\ref{fig_#1}}
\newcommand{\figs}[3]{Figures~\ref{fig_#1}(#2) and \ref{fig_#1}(#3)}
\newcommand{\tab}[1]{Table~\ref{tab_#1}}
\renewcommand{\S}{Section~}
\newcommand{\col}{\\[1ex](A color version of this figure is available in the online journal.)}

\newcommand{\avg}[1]{\overline{#1}}
\newcommand{\abs}[1]{|#1|}

\newcommand{\kms}{{\rm km\,s^{-1}}}
\newcommand{\msun}{{\rm M}_{\odot}}

\newcommand{\unitj}{\rm kpc\,\kms}

\newcommand{\ha}{H{\sc\,i}}
\newcommand{\hm}{H$_2$}

\newcommand{\Mb}{M_{\rm b}}
\newcommand{\Mh}{M_{\rm h}}
\newcommand{\Rh}{R_{\rm h}}
\newcommand{\Vh}{V_{\rm h}}
\newcommand{\Eh}{E_{\rm h}}
\newcommand{\Jh}{J_{\rm h}}
\newcommand{\jh}{j_{\rm h}}
\newcommand{\Ms}{M_{\ast}}
\newcommand{\Mg}{M_{\rm gas}}
\newcommand{\Mha}{M_{\rm H{\sc\,I}}}
\newcommand{\Mhm}{M_{\rm H_2}}
\newcommand{\Mdisk}{M_{\rm disk}}
\newcommand{\Mbulge}{M_{\rm bulge}}
\newcommand{\Jb}{J_{\rm b}}
\newcommand{\Js}{J_{\ast}}

\newcommand{\Jha}{J_{\rm H{\sc\,I}}}
\newcommand{\Jhm}{J_{\rm H_2}}
\newcommand{\jb}{j_{\rm b}}
\newcommand{\js}{j_{\ast}}
\newcommand{\jg}{j_{\rm gas}}
\newcommand{\jha}{j_{\rm H{\sc\,I}}}
\newcommand{\jhm}{j_{\rm H_2}}
\newcommand{\jdisk}{j_{\rm disk}}
\newcommand{\jbulge}{j_{\rm bulge}}

\newcommand{\fb}{f_{\rm M}}
\newcommand{\fj}{f_{\rm j}}

\newcommand{\Sha}{\Sigma_{\rm H{\sc\,I}}}
\newcommand{\Shm}{\Sigma_{\rm H_2}}
\newcommand{\Ss}{\Sigma_\ast}

\newcommand{\Ssfr}{\Sigma_{\rm SFR}}
\newcommand{\bt}{\beta}
\renewcommand{\lg}{{\rm\,lg\,}}

\newcommand{\Rs}{R_{\ast}}

\newcommand{\Rmax}{R_{\rm max}}
\newcommand{\kappan}{\kappa_0}
\newcommand{\Sn}{\Sigma_0}
\newcommand{\sigman}{\sigma_0}
\newcommand{\vflat}{V}

\begin{document}

\title{FUNDAMENTAL MASS--SPIN--MORPHOLOGY RELATION OF SPIRAL GALAXIES}

\author{D. Obreschkow$^{1,2}$}
\author{K. Glazebrook$^{2,3}$}
\affiliation{$^1$International Centre for Radio Astronomy Research (ICRAR), M468, University of Western Australia, 35 Stirling Hwy, Crawley, WA 6009, Australia}
\affiliation{$^2$ARC Centre of Excellence for All-sky Astrophysics (CAASTRO)}
\affiliation{$^3$Centre for Astrophysics and Supercomputing, Swinburne University of Technology, P.O. Box 218, Hawthorn, VIC 3122, Australia}
\date{\today}

\begin{abstract}
This work presents high-precision measurements of the specific baryon angular momentum $\jb$, contained in stars, atomic gas, and molecular gas, out to $\gtrsim\!10$ scale radii, in 16 nearby spiral galaxies of the THINGS sample. The accuracy of these measurements improves on existing studies by an order of magnitude, leading to the discovery of a strong correlation between the baryon mass $\Mb$, $\jb$, and the bulge mass fraction $\bt$, fitted by $\bt=-(0.34\pm0.03)\lg (\jb\Mb^{-1}/[\rm10^{-7}kpc\,km\,s^{-1}\,\msun^{-1}])-(0.04\pm0.01)$\!\! on the full sample range of $0\leq\bt\lesssim0.3$ and $10^9\msun<\Mb<10^{11}\msun$. The corresponding relation for the stellar quantities $\Ms$ and $\js$ is identical within the uncertainties. These $M$-$j$-$\bt$ relations likely originate from the proportionality between \smash{$j M^{-1}$} and the surface density of the disk that dictates its stability against (pseudo-)bulge formation. Using a cold dark matter model, we can approximately explain classical scaling relations, such as the fundamental plane of spiral galaxies, the Tully-Fisher relation, and the mass-size relation, in terms of the $M$-$j$(-$\bt$) relation. These results advocate the use of mass and angular momentum as the most fundamental quantities of spiral galaxies.
\end{abstract}

\maketitle


\section{Introduction}\label{section_introduction}

In galaxies, total mass $M$ and orbital angular momentum $J$ are fundamental concepts: they are conserved in isolated systems (\emph{invariance}), defined in any galaxy (\emph{universality}), and key to other properties (\emph{causality}). In fact, $M$ and $J$ collectively dictate the density normalization and radius of the galaxy-system \citep{Mo1998}. They thus set the disk pressure and associated physics, including phase transitions \citep{Blitz2006} and instabilities, which affect observables, such as luminosity and morphology. The key question to be answered here is how the primary morphological feature of disk galaxies, their bulge, depends on $M$ and $J$.

The fundamental nature of $M$ and $J$ motivates their use as primary parameters to describe galaxies \citep{Hernandez2006}. In doing so, it is common to remove the implicit mass scaling of $J$ by adopting the specific angular momentum $j\equiv J/M$. $M$ and $j$ are then independent in terms of basic units (mass versus length$^2/$time). In this work, $M$ and $j$ are indexed to distinguish between stars ($\ast$), neutral atomic gas (\ha), molecular gas (\hm), and all baryons in the galaxy ($b$). \ha\ and \hm\ include 36\% helium in addition to hydrogen, and the term `baryons' refers to the sum of stars, \ha, and \hm\ without including hot halo gas. The quantities $M$ and $j$ without subindices generally refer to either stars ($\Ms$ and $\js$) or baryons ($\Mb$ and $\jb$).

The first empirical investigation of galaxies in $M$-$j$ space was presented by \cite{Fall1983}. He used stellar masses $\Ms$ derived from total luminosities and approximate $\js$ to study a sample of 44 spiral (Sb-Sc) galaxies and a sample of 44 elliptical galaxies. In both samples, $\Ms$ and $\js$ were found to follow a relation $\js=q\Ms^\alpha$ with similar exponents $\alpha\approx2/3$, but a prefactor $q$  about 5-times lower in elliptical galaxies, indicating a significant loss of angular momentum in their formation history. The exponent $\alpha=2/3$ is a prediction of the cold dark matter (CDM) theory within some simplistic assumptions, while the factor $q$ depends more subtly on the baryon physics in ways sketched out by \cite{Romanowsky2012}. They revisit the $\Ms$-$\js$ relation of \cite{Fall1983} using a broader morphology range of 67 spiral (Sa--Sm) and 40 elliptical (E7--S0) galaxies. Their study represents the largest and most comprehensive investigation of galaxies in the $\Ms$-$\js$ plane to date. One of the prime results is that the Hubble sequence of galaxy morphologies is essentially a sequence of increasing angular momentum at any fixed mass -- confirming and refining an original suggestion by \cite{Sandage1970}.

A shortcoming in current measurements of angular momentum is that they do not include the contribution of gas and that stellar angular momenta $\Js$ are not actually measured by integrating $d\Js$ over the spatially and kinematically resolved galaxies. Instead, $\js$ is approximated as $\js=kv'r'$, where $k$ is a scalar parameter, $v'$ is a measure of the rotation velocity, and $r'$ a specific type of radius \cite[e.g.,~Equations (2) and (7) in][]{Romanowsky2012}. Requiring less data than a full measurement, this approximation can be applied to larger galaxy samples at the cost of introducing random and systematic errors in $\js$. More accurate measurements of $\js$ are technically difficult, because they require deep long-slit spectroscopy or kinematic maps with kpc resolution obtainable via integral field spectroscopy (IFS) -- a quickly rising 21st century technology \citep{Glazebrook2013}. For instance, SAGES Legacy Unifying Globulars and GalaxieS (SLUGGS, \citealp{Arnold2013}), a deep survey on the Keck/DEIMOS spectrograph, revealed converged measurements of $\js$ in six early-type galaxies \citep{Romanowsky2012}. Examples of IFS surveys enabling somewhat less accurate (since less deep) measurements of $\js$ include the ATLAS$^{\rm 3D}$ multi-wavelength IFS survey \citep{Cappellari2011}, the Calar Alto Legacy Integral Field Area Survey (CALIFA, \citealp{Sanchez2012}), the Mapping Nearby Galaxies at APO (MaNGA) survey, the survey with the Sydney Australian Astronomical Observatory Multi-object Integral Field Spectrograph (SAMI, \citealp{Croom2012}), and its proposed highly multiplexed successor (HECTOR, \citealp{Lawrence2012}).

On the theoretical side, both analytical models and numerical simulations are used to investigate the growth of $j$. Models assuming that the value of $j$ set by tidal torques during the protogalactic formation of structure \citep{Peebles1969,Doroshkevich1970,White1984} remains conserved during the formation of galaxies, except when large spheroids form, can reproduce the slope and zero-point of the $M$-$j$ relation \citep{Fall1983,Romanowsky2012}. Yet, until recently, hydrodynamic simulations indicated that $j$ is in fact not conserved, but significantly reduced by dynamical friction during the contraction of the gas. Consequently, simulated galaxies were systematically smaller and bulgier than observed ones \citep{Navarro2000,Stinson2010}. This `angular momentum crisis' hindered theoretical inferences from observed angular momenta. It now seems understood that the numerical loss of angular momentum was an artifact associated with insufficient spatial resolution and a lack of supernovae feedback that removes low-$j$ material from the galaxy centers \citep{Governato2010,Agertz2011,Guedes2011,Marinacci2014}. Simulations overcoming these challenges are about to reveal details of the joint growth of mass and angular momentum in galactic disks \citep[e.g.,][]{Brooks2011}. In parallel, semi-analytic models of millions of galaxies increasingly focus on angular momentum \citep{Benson2012} and have already uncovered the importance of the co-evolution of $M$ and $j$ in explaining the cosmic history of star formation \citep{Obreschkow2009c,Obreschkow2009d}.

\begin{figure}[t]
	\begin{tabular}{c}
		\put(-14,0){\includegraphics[width=1.06\columnwidth]{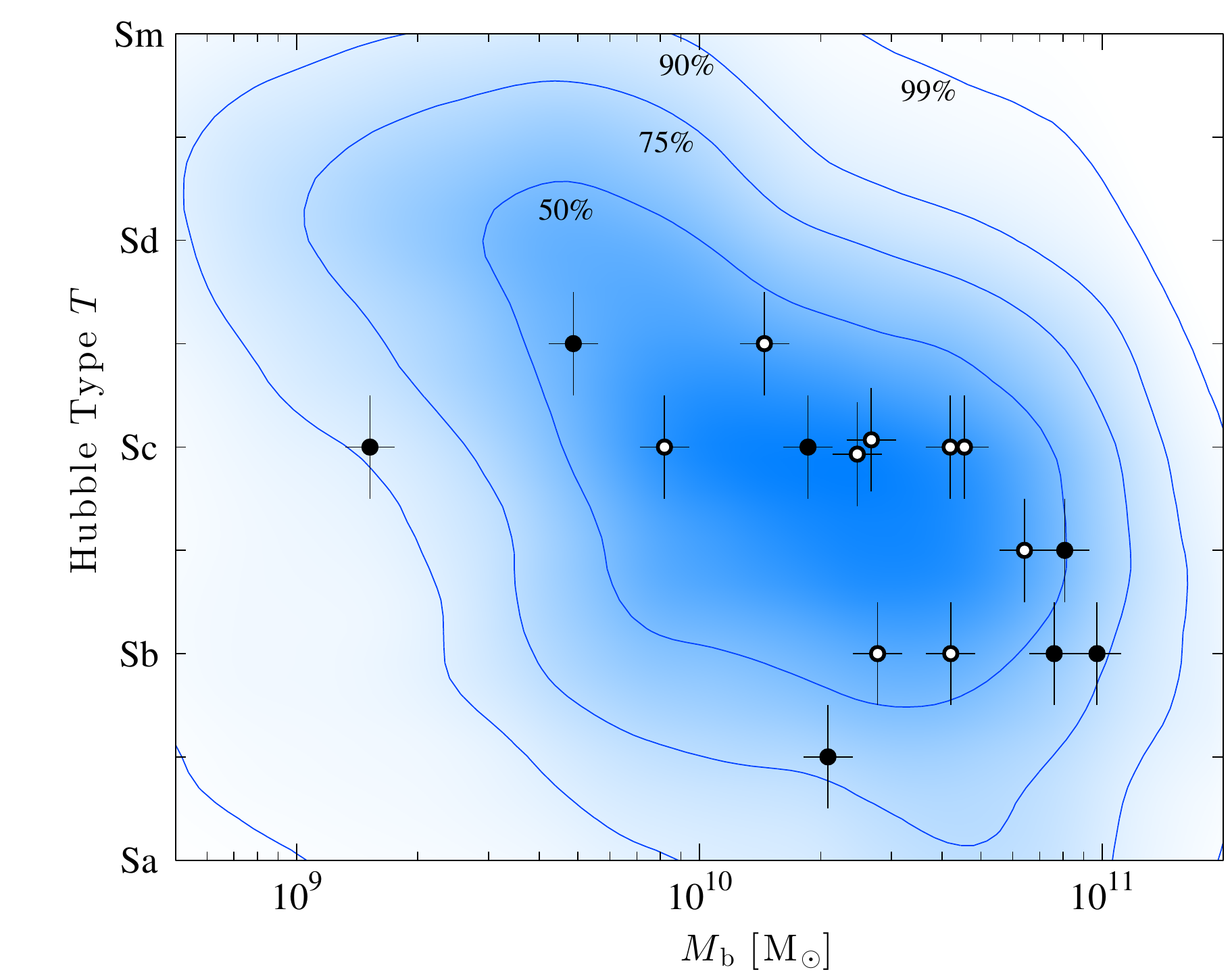}}
	\end{tabular}
	\caption{Representation of the 16 barred (open circles) and unbarred (filled circles) spiral galaxies in the ($\Mb$,$T$)-plane. Error bars are standard deviations of measurement uncertainties. The sample is compared against the 30-times larger reference sample of HIPASS galaxies with measured morphologies and baryon masses \citep{Meyer2008}. The distribution of this reference sample in the ($\Mb$,$T$)-plane, smoothed by a 2D-Gaussian Kernel matching the $(x,y)$-measurement uncertainties, is shown as a blue density field with contours containing the indicated fraction of galaxies. \col}
	\label{fig_sample}
\end{figure}

\emph{With IFS surveys flourishing and accurate simulations of angular momentum in large galaxy samples within reach, angular momentum is becoming a standard tool in galaxy evolution research.} This paper explores this new era with the aim to measure the $M$-$j$ relation in spiral galaxies and its dependence on morphology. Unprecedented precision is achieved using deep high-resolution ($<\rm kpc$) kinematic data available for 16 spiral galaxies of The \ha\ Nearby Galaxy Survey (THINGS, \citealp{Walter2008}). The observational accuracy of the resulting $j$-values exceeds existing studies by an order of magnitude, and for the first time the measurements also comprise the contributions of \ha\ and \hm\ in addition to stars. Using these data, the $M$-$j$-morphology relation of spiral galaxies turns out to be much tighter than previously known \citep{Romanowsky2012}.

Section \ref{section_data} introduces the sample of spiral galaxies and the method to compute their angular momenta. Section \ref{section_mj_relation} analyzes the $M$-$j$ relation (for stars and all baryons) and its dependence on the bulge mass fraction $\bt$ (often called B/T). A strong three-dimensional (3D) correlation is discovered and discussed in Sections \ref{section_discussion2D} and \ref{section_discussion3D}. Section \ref{section_conclusion} concludes the paper with a summary of the key results. An in-depth analysis of angular momentum contained in different gas phases, as well as additional scaling relations will be discussed in a sequel paper.


\section{Measurement of Angular Momentum}\label{section_data}
 
\subsection{Sample of Spiral Galaxies}
 
\begin{figure*}[t]
	\centering
	\begin{tabular}{cc}
		\begin{overpic}[width=\columnwidth]{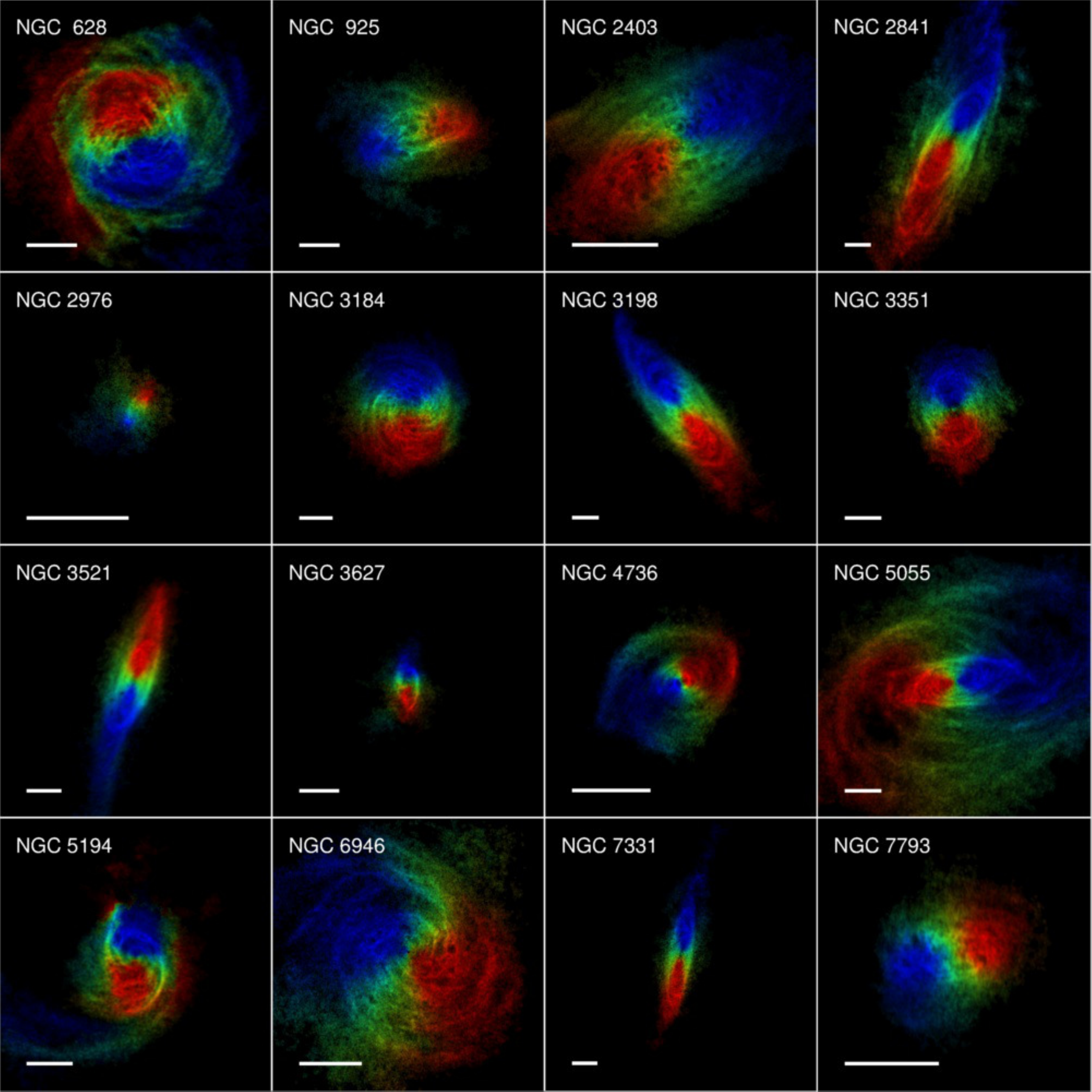}\end{overpic} & 
		\begin{overpic}[width=\columnwidth]{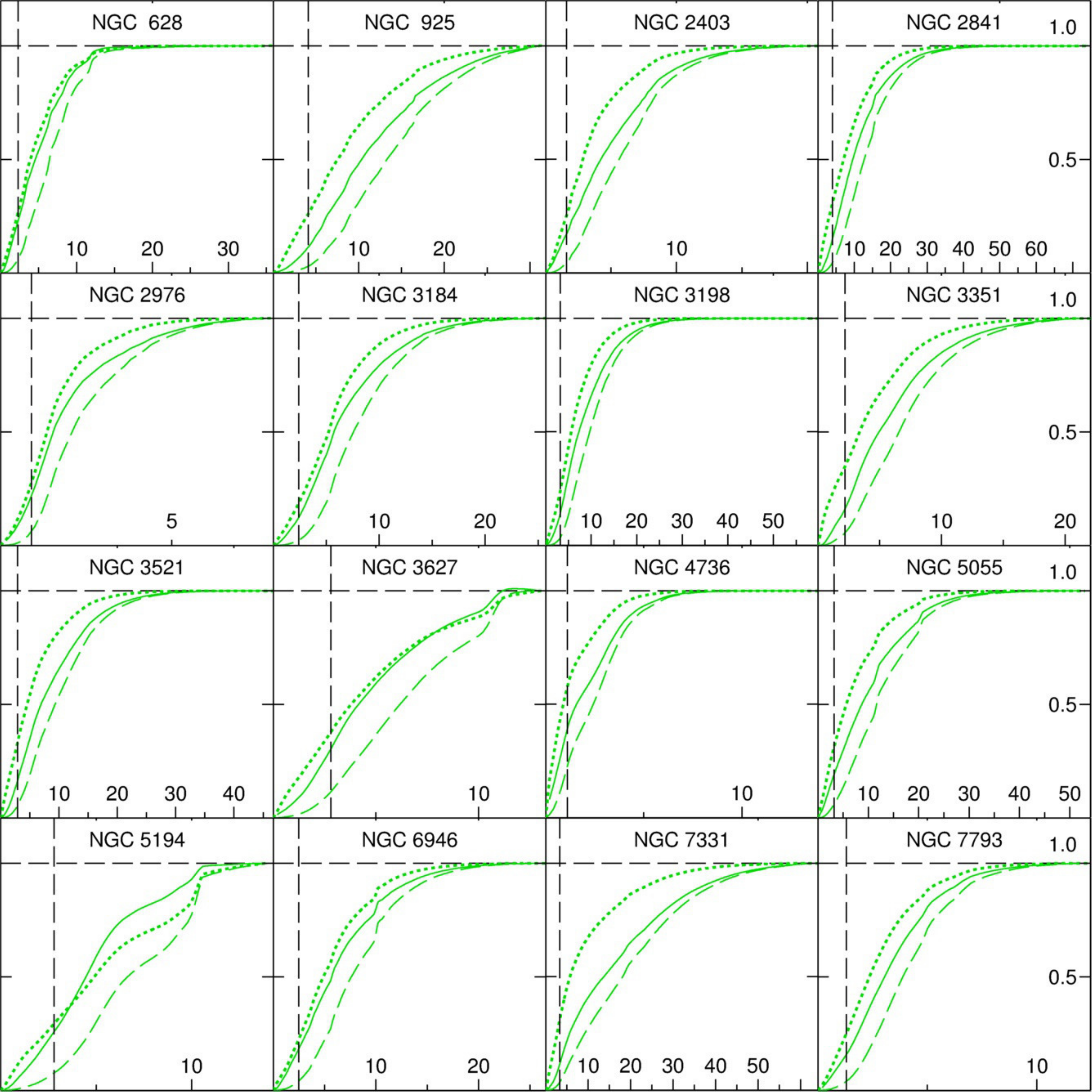}\end{overpic}
	\end{tabular}
	\caption{Left: combined \ha\ intensity map and color-coded \ha\ velocity map of the 16 spiral galaxies considered in this work. Colors range from red to blue for projected velocities from $-\vflat\sin i$ to $\vflat\sin i$, where $\vflat$ (\tab{objects}, from \citealp{Leroy2008}) is the asymptotic rotational velocity and $i$ is the galaxy inclination \citep{Leroy2008}. The white bars represent 10~kpc scales. Right: fraction of the stellar mass $\Ms$ (dotted), stellar angular momentum $\Js$ (dashed), and specific stellar angular momentum $\js=\Js/\Ms$ (solid), enclosed within a given radius. The vertical dashed lines represent the exponential scale length $\Rs$ (\tab{objects}, from \citealp{Leroy2008}).}
	\label{fig_things}
\end{figure*}

This study uses all 16 spiral\footnote{NGC 3077 is listed as an Sd spiral galaxy in \cite{Leroy2008}, but upon visual inspection this galaxy is removed, being an irregular object, in agreement with the interaction study of \cite{Walter2002} and the `I0 pec' classification in the NASA/IPAC Extragalactic Database (NED).} galaxies of the THINGS sample \citep{Walter2008}, for which stellar and cold gas surface densities have been published by \cite{Leroy2008}. This sample, shown in \fig{things} (left) and \tab{objects}, offers the highest quality data to date for a detailed measurement of $\js\equiv\Js/\Ms$, $\jha\equiv\Jha/\Mha$, and $\jhm\equiv\Jhm/\Mhm$ in spiral galaxies. The sample covers stellar masses from $10^9\msun$ to $8\cdot10^{10}\msun$ and Hubble types $T$ from Sab to Scd. \fig{sample} shows the 16 galaxies in the ($\Mb$,$T$)-plane on top of the distribution of galaxies in the \ha~Parkes All Sky Survey (HIPASS, \citealp{Barnes2001}) with resolved morphologies and $K$-band based stellar masses (494 galaxies, c.f.\ \citealp{Meyer2008}). This figure reveals that the 16 galaxies nicely represent the majority of spiral galaxies detected in a typical 21cm/optically limited survey.

\begin{table*}[t]
	\centering
	\normalsize
	\begin{tabular}{r|ccccc|ccccc|ccccc}
	\hline\hline
	NGC & Type & $\bt$ & $\Rs$ & $R_{\rm flat}$ & $\vflat$ & $\Mb$ & $\Ms$ & $\Mg$ & $\Mha$ & $\Mhm$ & $\jb$ & $\js$ & $\jg$ & $\jha$ & $\jhm$ \\ [0.5ex]
	          &           &       & $\rm kpc$ & $\rm kpc$ & $\rm km\,s^{-1}$ & \multicolumn{5}{c|}{$\lg (\msun)$} & \multicolumn{5}{c}{$\lg (\unitj)$} \\ [0.5ex]
	\hline
	628 & Sc   & 0.04 & 2.3 & 0.8 & 217 & 10.27 & 10.1 & 9.8 & 9.7 & 9.0 & 3.07 & 2.98 & 3.23 & 3.28 & 2.98 \\ [0.5ex]
925 & SBcd & 0.05 & 4.1 & 6.5 & 136 & 10.16 & 9.9 & 9.8 & 9.8 & 8.4 & 3.01 & 2.94 & 3.09 & 3.10 & 2.95 \\ [0.5ex]
2403 & SBc  & 0.02 & 1.6 & 1.7 & 134 & 9.91 & 9.7 & 9.5 & 9.5 & 7.3 & 2.85 & 2.62 & 3.04 & 3.08 & 2.61 \\ [0.5ex]
2841 & Sb   & 0.10 & 4.0 & 0.6 & 302 & 10.88 & 10.8 & 10.1 & 10.1 & 8.5 & 3.53 & 3.40 & 3.91 & 3.94 & 3.40 \\ [0.5ex]
2976 & Sc   & 0.00 & 0.9 & 1.2 & 92 & 9.18 & 9.1 & 8.4 & 8.3 & 7.8 & 2.05 & 2.03 & 2.11 & 2.21 & 1.93 \\ [0.5ex]
3184 & SBc  & 0.02 & 2.4 & 2.8 & 210 & 10.41 & 10.3 & 9.7 & 9.6 & 9.2 & 3.09 & 3.03 & 3.25 & 3.32 & 2.97 \\ [0.5ex]
3198 & SBc  & 0.03 & 3.2 & 2.8 & 150 & 10.41 & 10.1 & 10.1 & 10.1 & 8.8 & 3.24 & 2.97 & 3.45 & 3.49 & 3.02 \\ [0.5ex]
3351 & SBb  & 0.14 & 2.2 & 0.7 & 196 & 10.44 & 10.4 & 9.4 & 9.2 & 9.0 & 2.94 & 2.91 & 3.10 & 3.27 & 2.75 \\ [0.5ex]
3521 & SBbc & 0.16 & 2.9 & 1.4 & 227 & 10.81 & 10.7 & 10.1 & 10.0 & 9.6 & 3.14 & 3.06 & 3.38 & 3.46 & 2.98 \\ [0.5ex]
3627 & SBb  & 0.22 & 2.8 & 1.2 & 192 & 10.62 & 10.6 & 9.4 & 9.0 & 9.1 & 2.84 & 2.84 & 2.82 & 2.93 & 2.77 \\ [0.5ex]
4736 & Sab  & 0.32 & 1.1 & 0.2 & 156 & 10.32 & 10.3 & 9.0 & 8.7 & 8.6 & 2.37 & 2.34 & 2.63 & 2.86 & 2.36 \\ [0.5ex]
5055 & Sbc  & 0.17 & 3.2 & 0.7 & 192 & 10.91 & 10.8 & 10.2 & 10.1 & 9.7 & 3.30 & 3.18 & 3.59 & 3.69 & 3.08 \\ [0.5ex]
5194 & SBc  & 0.09 & 2.8 & 0.8 & 219 & 10.66 & 10.6 & 9.8 & 9.5 & 9.4 & 3.19 & 3.18 & 3.25 & 3.12 & 3.36 \\ [0.5ex]
6946 & SBc  & 0.10 & 2.5 & 1.4 & 186 & 10.62 & 10.5 & 10.0 & 9.8 & 9.6 & 3.06 & 3.02 & 3.17 & 3.32 & 2.91 \\ [0.5ex]
7331 & SAb  & 0.16 & 3.3 & 1.3 & 244 & 10.99 & 10.9 & 10.2 & 10.1 & 9.7 & 3.38 & 3.35 & 3.52 & 3.59 & 3.30 \\ [0.5ex]
7793 & Scd  & 0.01 & 1.3 & 1.5 & 115 & 9.69 & 9.5 & 9.2 & 9.1 & (8.6) & 2.49 & 2.43 & 2.61 & 2.66 & 2.40 \\ [0.5ex]

	\hline
	\hline
	\end{tabular}
	\caption{\upshape\raggedright Properties of the 16 spiral galaxies studied in this paper. The specific angular momenta $j$ were calculated as described in \S\ref{subsection_j_measurement}. The bulge mass fractions $\bt$ were computed as explained in \S\ref{subsection_data} and illustrated in \fig{disk_bulge}. All other values have been copied from Table~4 in \cite{Leroy2008}, inferring $\Mhm$ of NGC~7793 from its SFR as described in \S\ref{subsection_data} and using $\Mg=\Mha+\Mhm$ and $\Mb=\Ms+\Mg$. As explained by \citeauthor{Leroy2008}, the scale radius $\Rs$ and the rotation parameters $R_{\rm flat}$ and $\vflat$ represent fits to the stellar surface density $\Ss(r)\propto\exp(-r/\Rs)$ and \ha\ velocity profile $v(r)=\vflat[1-\exp(-r/R_{\rm flat})]$. Measurement uncertainties are not shown in this table, but they are plotted as error bars in the figures of Sections \ref{section_introduction} to \ref{section_discussion3D} and accounted for in all results.}
	\label{tab_objects}
\end{table*}

\subsection{Primary Data}\label{subsection_data}

The data collected by \cite{Leroy2008} comprises multi-wavelength maps from different surveys: kinematic \ha\ maps at a mean resolution of $11''$ ($\sim400\rm~pc$) and $5~\kms$ from The \ha\ Nearby Galaxy Survey (THINGS; \citealp{Walter2008}), far-ultraviolet (FUV) maps of $5.6''$ resolution from the space-based Galaxy Evolution Explorer (GALEX) Nearby Galaxies Survey \citep{GilDePaz2007}, $24\rm~\mu m$ and $3.6\rm~\mu m$ infrared (IR) data with a resolution of $\leq6''$ from the space-based Spitzer Infrared Nearby Galaxies Survey (SINGS; \citealp{Kennicutt2003}), CO$(1\rightarrow0)$ maps of $7''$ resolution from the Berkeley-Illinois-Maryland Association (BIMA) Survey of Nearby Galaxies (BIMA SONG; \citealp{Helfer2003}), and CO$(2\rightarrow1)$ maps of $11''$ resolution from the HERA CO Line Extragalactic Survey (HERACLES; \citealp{Leroy2009}).

From these data \cite[][c.f.~Appendices A-E therein]{Leroy2008} computed radial surface density profiles $\Sigma(r)$ as a function of radius $r$ at a resolution of $\sim\!400\rm~kpc$, degrading the raw resolution where necessary. Atomic gas densities $\Sha$ were computed from the integrated intensity maps of the 21~cm emission line. Molecular gas densities $\Shm$ were estimated from the CO$(2\rightarrow1)$ maps, except in the case of NGC 3627 and NGC 5194, where CO$(1\rightarrow0)$ maps were used instead. These estimates rely on a constant CO-to-\hm\ conversion factor $X_{\rm CO(1\rightarrow0)}=2\cdot10^{20}\rm~cm^{-2}\rm~(K\,km\,s^{-1})^{-1}$ with an additional correction of 1.36 to include helium, and a fixed line ratio $I_{\rm CO(2\rightarrow1)}=0.8I_{\rm CO(1\rightarrow0)}$. Stellar mass densities $\Ss$ were inferred from the 3.6~$\rm \mu m$ continuum maps. These maps were first reduced to median radial profiles to minimize the contribution of hot dust and polycyclic aromatic hydrocarbon (PAH) emission near star-forming regions. The median $3.6~\rm \mu m$ profiles were then converted to $\Ss(r)$ by adopting an empirical $K$-to-3.6~$\rm \mu m$ calibration and a constant $K$-band mass-to-light ratio of $\Upsilon_\ast^K=0.5~\msun/{\rm L}_{\odot,K}$, neglecting local variations of a factor $\sim\!2$ between young and old stellar populations. Star formation rate (SFR) surface densities $\Ssfr$, used to complete missing \hm\ data (see below), were derived from a combination of FUV and far-IR (FIR) 24~$\rm \mu m$ continuum maps to capture both directly visible and dust-obscured star formation (Appendix D of \citeauthor{Leroy2008}).

This paper uses the surface density profiles published by \cite{Leroy2008} up to the following variations. First, surface densities $\Sha(r)$ were re-derived from the \ha\ intensity maps \citep{Walter2008}, since the $\Sha(r)$ published by \citeauthor{Leroy2008}\ are restricted to $\geq1~\msun\rm pc^{-2}$. From the \ha\ maps most $\Sha(r)$ can be measured down to about $10^{-2}~\msun\rm pc^{-2}$. Using these extended data, it turns out that limiting $\Sha$ to $\geq1~\msun\rm pc^{-2}$ decreases $\Jha$ and $\jha$ by about 20\% and 10\%, respectively. These percentages improve to $1\%$ and $0.1\%$ if densities down to $10^{-1}~\msun\rm pc^{-2}$ are included, thus motivating the use of the full data. Second, where CO-based \hm\ surface densities are missing, they are estimated using an inverted star-formation law $\Shm=t_{\rm H_2}~\Ssfr$, where $t_{\rm H_2}=1.9\cdot10^9\rm~yr$ is the effective \hm\ depletion time found by \cite{Leroy2008}. This method is used to infer the total \hm\ mass of NGC 7793, the full functions $\Shm(r)$ of NGC 628/925/2403/2841/7793, as well as large-$r$ parts of $\Shm(r)$ in the other galaxies. Third, the densities $\Ss(r)$ and $\Shm(r)$ are extrapolated beyond the maximal radii $\Rmax$, to which they were measured or estimated. The extrapolations use an exponential profile $\Sn\exp(-r/R )$, with parameters $\Sn$ and $R$ fitted to the data on the range $r\in[\Rmax/2,\Rmax]$. The extrapolated parts are shown as dashed lines in \fig{profiles} (left). We emphasize that completing \hm\ data from SFRs and extrapolating $r$ beyond $\Rmax$ has no effect on the conclusions of this paper. This post-processing only affects $\jhm$, $\js$, and $\jb$ by $\sim10\%$ allowing these values to converge to the $1\%$ level (see \S\ref{subsection_j_measurement}).

To study correlations between angular momentum and galaxy morphology, the latter is quantified using the stellar mass fraction $\bt$ in the bulge. In this paper, `bulge' generically refers to any central stellar over-density without further specifying the nature of this component. In the present sample, these bulges are mostly flattened pseudo-bulges \citep{Kormendy2008}, nine of which include a bar component. For each galaxy, $\bt$ is calculated by fitting $\Ss(r)$ with a model composed of an exponential function for the disk and a S{\'e}rsic profile \citep{Sersic1963} for the bulge, as described in Appendix \ref{appendix_disk_bulge}. The resulting values of $\bt$ are listed in \tab{objects}. The standard errors inferred from resampling are about 0.02.

\subsection{Precision Measurement of Angular Momentum}\label{subsection_j_measurement}

In the approximation of a flat galaxy with circular orbits, the norm of the angular momentum relative to the center of gravity can be written as
\be\label{eq_J_general}
	J = \left|\int dM~\mathbf{r}\times\mathbf{v}\,\right| = 2\pi \int_0^\infty dr\,r^2\,\Sigma(r)\,v(r),
\ee
where $d M$ is the mass element, $\mathbf{r}$ is the position vector from the center of gravity, $\mathbf{v}$ is the velocity vector, $v(r)$ is the norm of $\mathbf{v}$ at $r=\abs{\mathbf{r}}$, and $\Sigma(r)$ is the azimuthally averaged mass surface density of the considered baryonic component. The specific angular momentum is
\be\label{eq_j}
	j \equiv \frac{J}{M} = \frac{\int_0^\infty dr~r^2~\Sigma(r)~v(r)}{\int_0^\infty dr~r~\Sigma(r)}.
\ee
Computing $J$ and $j$ from axially averaged density and velocity profiles allows the outskirts (to $r\approx14\Rs$) with low pixel signal-to-noise to be reliably included, but the use of axially averaged surface densities $\Sigma(r)$ does not, in fact, assume or require $\Sigma$ to be axially symmetric.

The integral of \eq{j} is evaluated numerically, while correcting for the inclination of the galaxy as detailed in Appendix \ref{appendix_J}. The integrals are evaluated out to the maximal observed \ha\ radius $R_{\rm HI,max}\approx14\Rs$. The only exception is NGC~5194 -- the Whirlpool Galaxy -- where the upper bound of the integral is restricted to $14~\rm kpc$ to suppress the contributions of the interacting close companion NGC~5195 and associated stripped material. \eq{j} is applied to the different baryonic surface densities, resulting in distinct values of $j_{\rm X}\equiv J_{\rm X}/M_{\rm X}$ for all the baryons ($\jb$), stars ($\js$), atomic gas ($\jha$), molecular gas ($\jhm$), and atomic and molecular gas together ($\jg$). These values are listed in \tab{objects}.

\begin{figure*}[t]
	\centering
	\begin{tabular}{cc}
		\begin{overpic}[width=8.6cm]{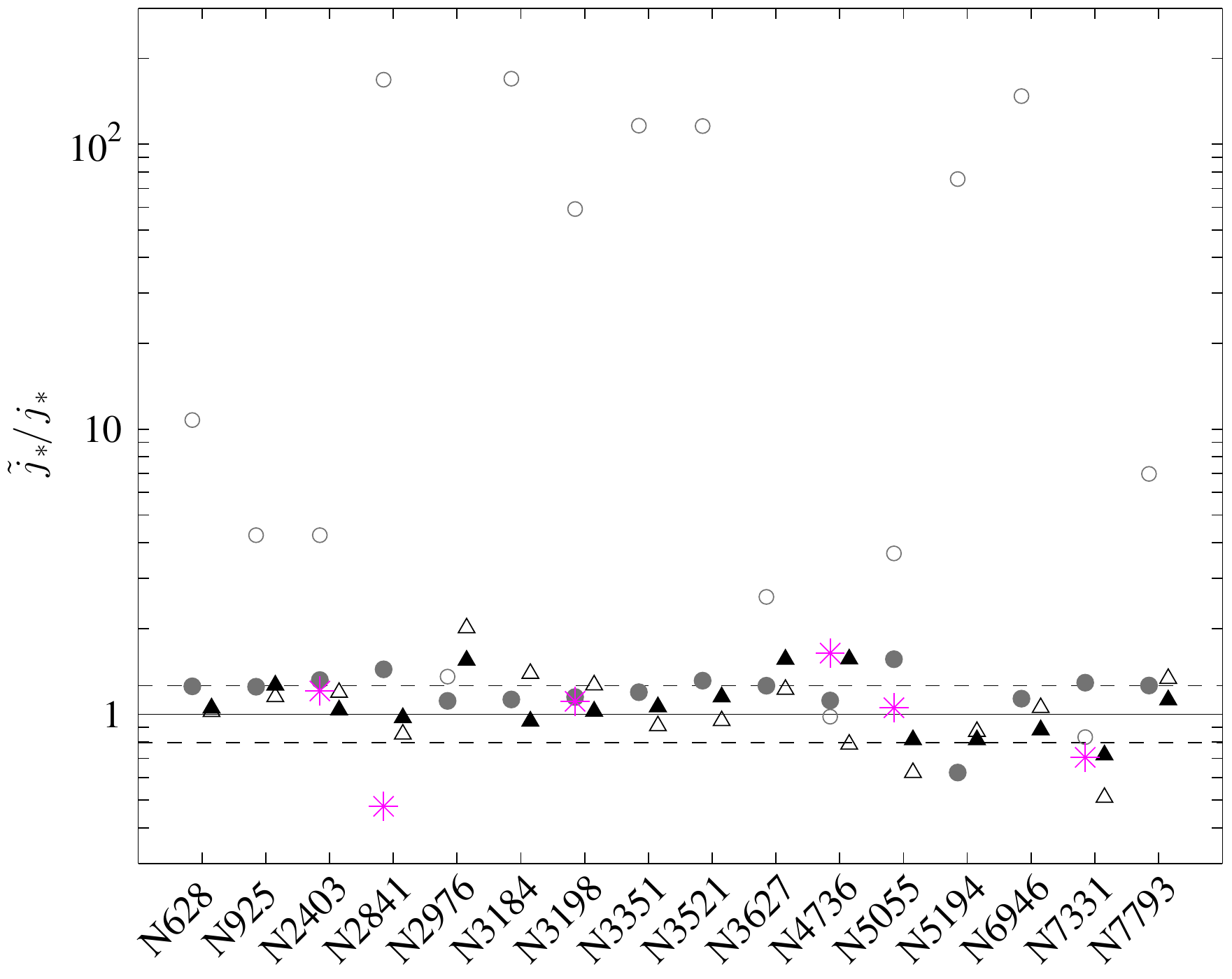}\put(-2.5,75.5){\normalsize\textbf{(a)}}\end{overpic} & 
		\begin{overpic}[width=8.75cm]{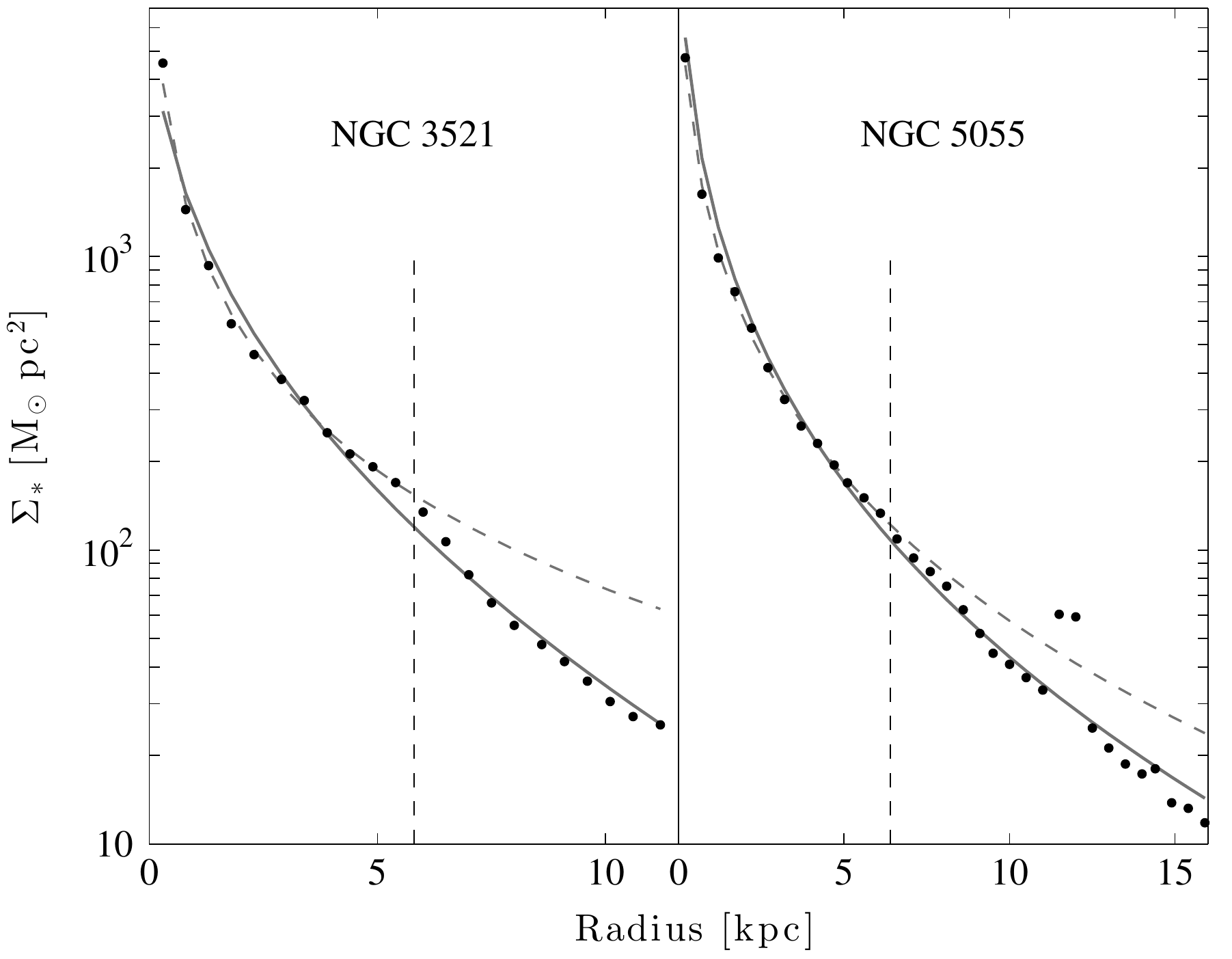}\put(0,74){\normalsize\textbf{(b)}}\end{overpic} \\ [4.0ex]
	\end{tabular}\vspace{-0.2cm}
	\caption{Assessment of approximate methods to measure the specific stellar angular momentum $\js$. (a) Approximate values $\tilde\js$, normalized by the full measurements $\js$. Triangles represent the exponential disk model of \eq{japprox1}, fitted to the full profiles $\Sigma(r)$ (filled triangles) and only to $r\leq2\Rs$ (open triangles). Circles represent the S{\'e}rsic approximation of \eq{japprox2}, fitted to the full profiles $\Sigma(r)$ (filled circles) and only to $r\leq2\Rs$ (open circles). Pink stars represent the estimates of $\js$ from \citeauthor{Romanowsky2012} for the six galaxies also contained in the present sample. (b) Functions $\Ss(r)$ for two selected galaxies. Black dots denote the measurements \citep{Leroy2008}, while lines show the S{\'e}rsic fits to the whole data (solid lines) and to $r\leq2\Rs$ (dashed lines). Those two types of fits correspond to the filled and open circles in panel~(a), respectively. Vertical dashed lines denote the limit $r=2\Rs$.\col\\}
	\label{fig_japprox}
\end{figure*}

All measurements of $j$ assume that the baryonic material orbits at the circular velocity $v(r)$ of the \ha\ gas. This assumption of co-rotation between \ha, \hm, and stars is justified in the rotation supported parts of the galaxy, where $v(r)$ is dictated by the local gravitational force. In the dispersion supported stellar bulge, however, the stellar rotational velocity is generally smaller than that of the \ha\ disk. For example, in the Andromeda galaxy (M31), the bulge rotation at $r\approx0-15\rm~kpc$ (about $50~\kms$, \citealp{Dorman2012}) is five times smaller than the disk rotation inferred from \ha\ (about $250~\kms$, \citealp{Unwin1983}). One might thus suspect that the values of $\js$ presented here over-estimate the real values. This effect is nonetheless negligible in late-type galaxies. In fact, even when using \ha\ velocities $v(r)$ for all stars, the stellar angular momentum $\Js$ of the bulge (according to the bulge-disk decomposition of \eq{disk_bulge}) only accounts for $0.3\%$ of the total $\Js$ on average. The most extreme bulge contributions are found in NGC~4736 (1.3\%), NGC~3627 (0.6\%), and NGC~5055 (0.6\%). Thus, the contribution of the \emph{angular momentum} of the stellar bulge to $\js$ is smaller than the statistical measurement uncertainties of a few percent for $\js$ (see hereafter). Bulges nonetheless affect $\js\equiv\Js/\Ms$ through their \emph{mass}, which takes values up to 0.32 of the total stellar mass $\Ms$ in the present sample.

How accurate are the measurements of $j$? Let us first discuss statistical uncertainties. By repeating the computations of $j$ via \eq{j} with random Gaussian variations of $\Sigma(r)$ matching their $r$-dependent measurement uncertainties, it turns out that such uncertainties affect $j$ by less than $0.1\%$. These errors are negligible relative to those associated and $v(r)$. In computing $v(r)$ via \eq{deprojection_v}, the inclination-dependent deprojection factor $C(\varphi,i)$ introduces an uncertainty in the normalisation of $v(r)$ of 2\%-4\% for the given inclination uncertainties. A second order uncertainty can result from non-circular orbits, since a non-circular velocity component $v_\perp$, perpendicular to the circular orbit, can perturb the measurement of the circular component $v(r)$. To estimate the magnitude of this effect, we generated $10^5$ mock galaxies, inclined at $51^\circ$ (the average inclination of the present sample), with constant $v(r)=V_0=200~\kms$, and a non-circular dipole component of amplitude $v_\perp=10~\kms$, typical for spiral galaxies \citep[e.g.][]{Beauvais1999}. For every mock galaxy, the orientation of the non-circular component in the plane perpendicular to the circular motion was chosen randomly. For each mock galaxy, we then recovered a circular velocity $V$ from the line-of-sight component $v_z$ (see \fig{deprojection}), assuming only circular orbits. The resulting values $V$ are centred on $V_0$, but scattered with a standard deviation of $4~\kms$ (2\%). Hence, the statistical error introduced when assuming circular orbits in the presence of a realistic non-circular component is approximately 2\%. Another source of statistical uncertainty is associated with the finite maximal observing radii $\Rmax$. In fact, due to the $r^2$ term in the angular momentum integral, non-detected low-density material in the outer ($r>\Rmax$) regions contributes more significantly to $J$ than to $M$. Thus the question, to what extent $j$ converges within $r\leq\Rmax$, requires careful examination. \fig{things} (right) shows the cumulative specific angular momenta $j(r)\equiv\int_0^r dr'\,r^2\Sigma(r')v(r')/\int_0^r dr'\,r\Sigma(r')$ of stars (for \ha\ and \hm\ see \fig{profiles}, right). To assess how well these functions have converged, a model for their extrapolation beyond $\Rmax$ is needed. Upon assuming an exponential disk $\Sigma(r)\propto\exp(-r/R )$ rotating at a constant circular velocity $V$, $j(r)$ becomes
\be\label{eq_jR}
\vspace{-0.1cm}
	j(r) = \left[2+\frac{(r/R )^2}{1+r/R -\exp(r/R )}\right]R \,V.
\ee
Explicit fits of \eq{jR} to the measured $j(r)$ predict that the measured $j$ have converged at the 1\% level for $\jb$, $\js$, and $\jhm$, and at the 10\% level for $\jha$ and $\jg$. Details and exceptions are given in Appendix \ref{appendix_J}.

The measurements of $j$ might also be subject to systematic errors. Errors in light-to-mass conversions, i.e., luminosity-to-stellar mass and CO-to-\hm, equally affect $J$ and $M$, thus canceling out in $j$. Only variations of these conversions within a galaxy can affect $j$. This might be significant for the CO-to-\hm\ conversion, which can vary along $r$ due to a metallicity gradient. A few available measurements for NGC~5194 \citep{Arimoto1996} suggest that the \hm/CO ratio increases by a factor $\sim\!2$ on two exponential scale radii. Accounting for this variation increases $\jhm$, $\jg$, and $\jb$ in NGC~5194 by about 20\%, 10\%, and 2\%, respectively. Similar changes might apply to other galaxies in the sample. However, since the CO-to-\hm\ conversion remains uncertain \citep{Obreschkow2009a}, this paper maintains the constant value of \cite{Leroy2008}. Other errors can result from a breakdown of the flat disk model in the case of disturbed or warped galaxies, but in the present sample such effects are negligible based on visual inspection. 
Distance errors affect $j$ linearly. The 16 galaxies considered here have Hubble flow distances (Table 1 in \citealp{Walter2008}) on the order of 10~Mpc with expected uncertainties around 5\% that are partially correlated.

In summary, the specific angular momenta have statistical uncertainties of a few percent (3\%-5\%) for $\jb$, $\js$, and $\jhm$, and $\sim\!10\%$ for $\jha$ and $\jg$. Potential systematic uncertainties are estimated to about 10\%.

\subsection{Comparison Against Approximate Measurements}\label{subsection_approximations}

Most measurements of angular momentum in the literature do not have detailed kinematic maps at their disposal. They therefore resort to approximations of $j$ based on global measurements. In this section, we compare typical approximations of $\js$, labeled as $\tilde\js$, against our precision measurements $\js$. Since the typical deviations between $\tilde\js$ and $\js$ turned out to be much larger than the few percent statistical uncertainties of $\js$, the latter can be considered as exact in this comparison.

The most common approximation of $\js$, already used by \cite{Fall1983}, relies on the flat, exponential disk model of \eq{jR}. In the limit of $r\rightarrow\infty$ this equation reduces to \citep[e.g., Equation (7) in][]{Mo1998},
\be\label{eq_japprox1}
	\tilde\js = 2\Rs V,
\ee
requiring only the exponential scale radius $\Rs$ of the stellar disk and the (constant) circular velocity $V$. Those two parameters can be estimated from other measurements, for instance $\Rs\approx0.6r_{\rm e}\approx0.3r_{25}$, where $r_{\rm e}$ is the `effective radius' containing half the light and $r_{25}$ is the `isophotal radius' with a $B$-band surface brightness of $25\rm~mag~arcsec^{-2}$. The velocity $V$ can be estimated from the total \ha\ linewidth or from optical linewithds at the radius $r_{25}$ (or beyond), corrected for turbulence and galaxy inclination. \fig{japprox}a (triangles) shows the values $\tilde\js$ given by \eq{japprox1}, normalized to the reference values $\js$. Filled triangles use $\Rs$ and $V$ derived from the full stellar surface densities $\Ss(r)$ and deprojected velocity profiles $v(r)$; they are the values $\Rs$ and $\vflat$ (\tab{objects}) adopted from \cite{Leroy2008}. Open triangles use approximate scale radii, fitted only to $r\leq2\Rs$. In general, this approximation based on the exponential disk model provides remarkably good results. The Root-Mean-Square (RMS) error of $\tilde\js$ for all 16 galaxies is about 30\% (0.10~dex). Using only data within $r\leq2\Rs$, this RMS error increases to 40\% (0.14~dex).

\begin{figure}[t]
	\includegraphics[width=\columnwidth]{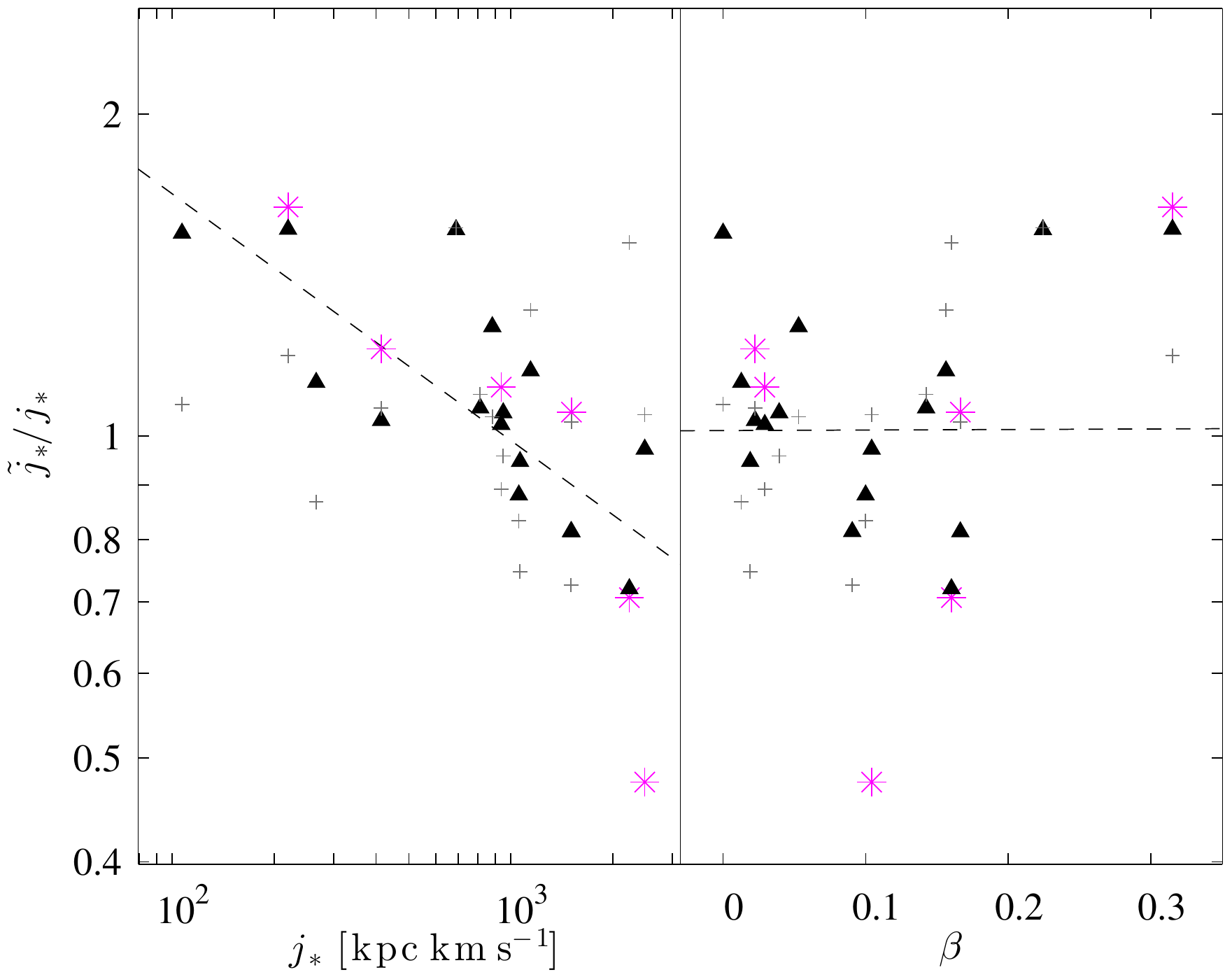}
	\caption{The approximate specific angular momenta $\tilde\js$, calculated with \eq{japprox1} (triangles) and adopted from \cite{Romanowsky2012} (pink stars), deviate \emph{systematically } from $\js$. This dependence is fitted by \eq{fit_j_from_japprox}, shown as dashed line in the left panel. This relation can be explained by a systematic variation of the stellar surface density and rotation curve with galaxy mass. Accounting for these variations removes the correlation between $\js$ and $\tilde\js/\js$ (crosses). No significant correlation is detected between $\tilde\js/\js$ and $\bt$.\col}
	\label{fig_japprox_systematic}
\end{figure}

\begin{figure}[t]
 	\centering
	\includegraphics[width=\columnwidth]{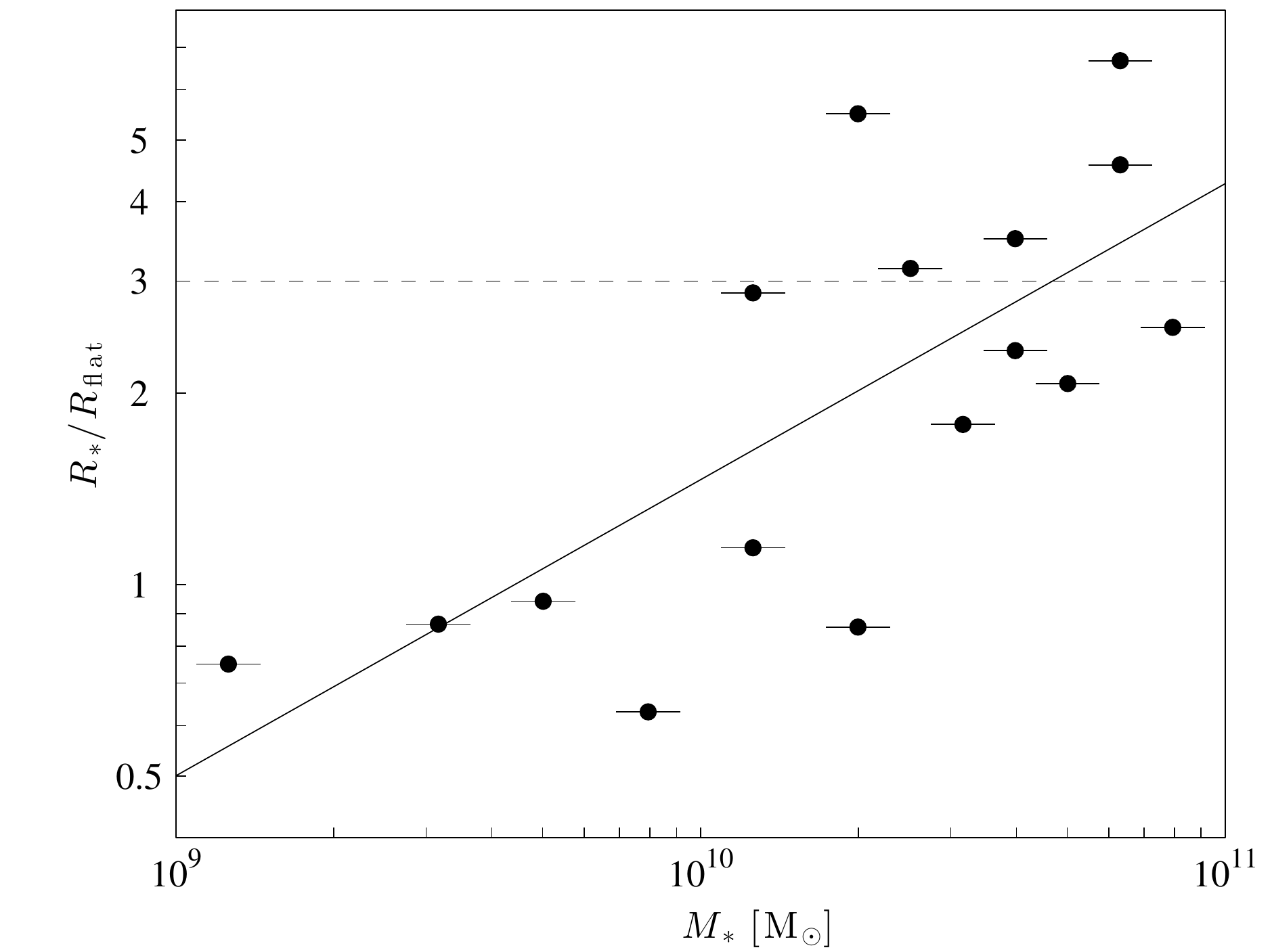}
	\caption{Dependence of the disk concentration, parameterized by $\Rs/R_{\rm flat}$, on stellar mass $\Ms$ (values from \citealp{Leroy2008}, shown in \tab{objects} of this paper). The solid line represents a standard linear regression, whereas the dashed line indicates the value $\Rs/R_{\rm flat}=3$, typical for Milky Way-sized spiral galaxies.}
	\label{fig_rstar_rflat}
\end{figure}

Another approximation, introduced by \cite{Romanowsky2012}, builds on the flat disk model with a surface density described by the S{\'e}rsic profile $\Sigma(r)\propto\exp[-b_n(r/r_{\rm e})^{1/n}]$ with free parameters $r_{\rm e}$ and $n$ ($n=1$ for exponential disk, $n=4$ for a de Vaucouleurs profile). The factor $b_n\approx2n-1/3+0.009876/n$ ensures that $r_e$ is the effective radius, $\int_0^{r_{\rm e}}dr\,r\,\Sigma(r)=0.5\int_0^\infty dr\,r\,\Sigma(r)$. \citeauthor{Romanowsky2012} find that $\js$ is approximated by
\be\label{eq_japprox2}
	\tilde\js=k_n v_{\rm s} r_{\rm e},
\ee
where $v_{\rm s}$ is the deprojected rotation velocity measured at a radius $2r_{\rm e}$ and $k_n\approx1.15+0.029n+0.062n^2$. To test this approximation the 16 galaxies of this work were fitted with single S{\'e}rsic functions, once using the whole profiles $\Ss(r)$, once artificially restricting them to radii $r\leq2\Rs$, where $\Rs$ is again the exponential scale radius given by \cite{Leroy2008}. The velocities $v_{\rm s}=v(2r_{\rm e})$ are then taken as the average of the de-projected \ha\ velocity $v(r)$ between $1.9r_{\rm e}$ and $2.1r_{\rm e}$. The resulting approximations $\tilde\js$ are shown in \fig{japprox}a (circles). If the S{\'e}rsic functions are fitted to the full data (filled circles in \fig{japprox}a), that is roughly within $r\leq14\Rs$, the RMS error is about 30\% (0.11~dex), comparable to the exponential disk model. However, when fitting only within $r\leq2\Rs$ (open circles), the RMS error heavily increases to 2500\% (1.4~dex), with $\tilde\js$ being systematically larger than $\js$. This large error can be traced back to the fact that S{\'e}rsic functions fitted to the inner ($r\leq2\Rs$), bulgier part of the galaxy systematically overestimate the surface density at larger radii by overestimating the index $n$, as illustrated in \fig{japprox}b. In conclusion, the S{\'e}rsic approximation of \eq{japprox2} is much more prone to errors than the exponential disk approximation of \eq{japprox1}.

\cite{Romanowsky2012} do not, in fact, use the S{\'e}rsic approximation of \eq{japprox2} to estimate $\js$ of spiral galaxies, since they also find the S{\'e}rsic fits to be too uncertain. Instead, they adopt a more robust approach that separates the galaxy into an exponential disk and a smaller `classical' bulge with a de Vaucouleurs profile (fixed S{\'e}rsic index $n=4$). For both components $\js$ is approximated separately and then recombined. Six of the Romanowsky galaxies are also in the present sample. Their values $\tilde\js$, plotted in \fig{japprox} (pink stars), yield an RMS error of about 50\% (0.17~dex).

A serious concern is that the errors of the approximations $\tilde\js$ correlate significantly with $\js$ (albeit not with $\bt$), as shown in \fig{japprox_systematic}. This correlation applies both to the $\tilde\js$ calculated via \eq{japprox1} (triangles in \fig{japprox_systematic}) and to those determined by \cite{Romanowsky2012} (pink stars). The correlation is best fitted by the dashed line in \fig{japprox_systematic}(a), which can be rewritten as
\be\label{eq_fit_j_from_japprox}
	\left[\frac{\js}{10^3\unitj}\right]\approx1.01\left[\frac{\tilde\js}{10^3\unitj}\right]^{ 1.3}.
\ee
This non-linearity between $\js$ and $\tilde\js$ is traceable to two features. Firstly, the stellar surface density $\Ss(r)$ systematically deviates from an exponential in such a way that the fraction $f_J$ of stellar angular momentum outside the half-light radius $r_{\rm e}$ increases with mass. This fraction ranges from about $f_{J}\approx75\%$ at $\Ms=10^9\msun$ to $f_{J}\approx85\%$ at $\Ms=10^{11}\msun$. To account for the high values of $f_J$ and their variability, the scale radius $R_\ast$ used in \eq{japprox1} can be fitted on $r>r_{\rm e}$ rather than on the whole disk. When doing so, the correlation between $\lg(\tilde\js/\js)$ and $\lg(\js)$ is reduced by 60\%. The remaining 40\% are traceable to a systematic variation of the rotation curves $v(r)$ with mass. This can be seen by looking at the fits \citep{Boissier2003}
\be\label{eq_vfit}
	v(r)\approx \vflat\left[1-\exp\left(-\frac{r}{R_{\rm flat}}\right)\right],
\ee
performed by \cite{Leroy2008}; their best-fitting parameters $\vflat$ and $R_{\rm flat}$ are listed in \tab{objects}. The ratio between Leroy's stellar scale radius $\Rs$ and $R_{\rm flat}$ is found to increase with $\Ms$, roughly by a factor 3 per dex in $\Ms$, as shown in \fig{rstar_rflat} (see also \citealp{DeBlok2008}). Thus, the normalized rotation curves $v(r/\Rs)/\vflat$ increase faster in more massive galaxies -- an effect that is also seen in radial variations of the Tully-Fisher relation in larger galaxy samples \citep{Yegorova2007}. We can account for this effect by convolving \eq{vfit} with an exponential surface density $\Ss(r)\propto\exp(-r/R_\ast)$ in \eq{j}. This solves to
\be\label{eq_japprox3}
	\tilde\js = 2\Rs\vflat\frac{(\Rs+R_{\rm flat})^3-R_{\rm flat}^3}{(\Rs+R_{\rm flat})^3}.
\ee
Using \eq{japprox3} with $R_\ast$ fitted to $\Ss(r)$ on $r>r_{\rm e}$ completely removes the correlation between $\js$ and $\tilde\js/\js$ (crosses in \fig{japprox_systematic}(a)). In conclusion, $\tilde\js$ approximated by \eq{japprox1} and \cite{Romanowsky2012} is offset from $\js$ via \eq{fit_j_from_japprox} due to a systematic mass dependence of the disk shape and rotation curve.


\begin{figure*}[t]
	\centering
	\begin{tabular}{cc}
		\begin{overpic}[width=\columnwidth]{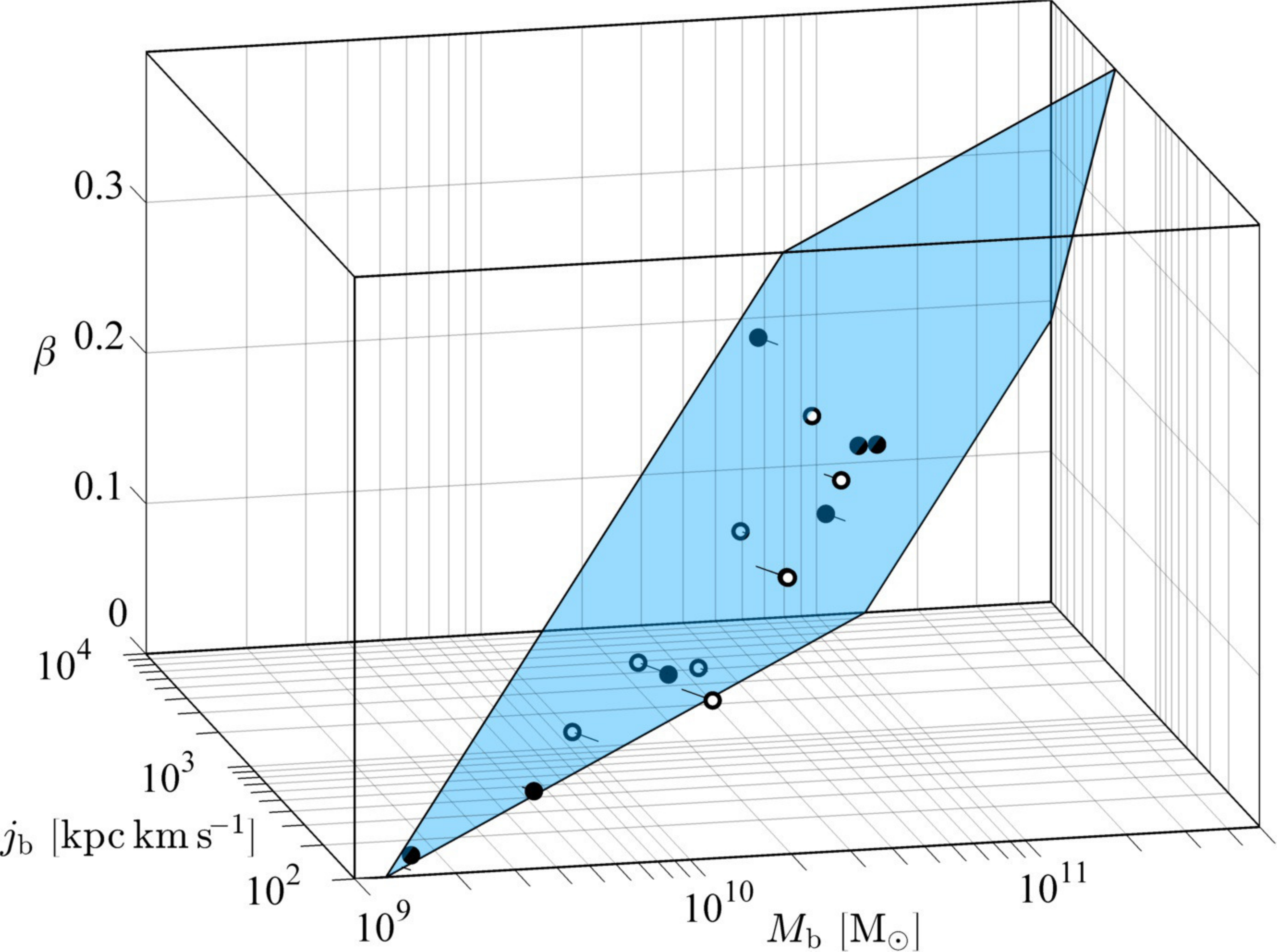}\put(0,71){\normalsize\textbf{(a)}}\end{overpic} & 
		\begin{overpic}[width=\columnwidth]{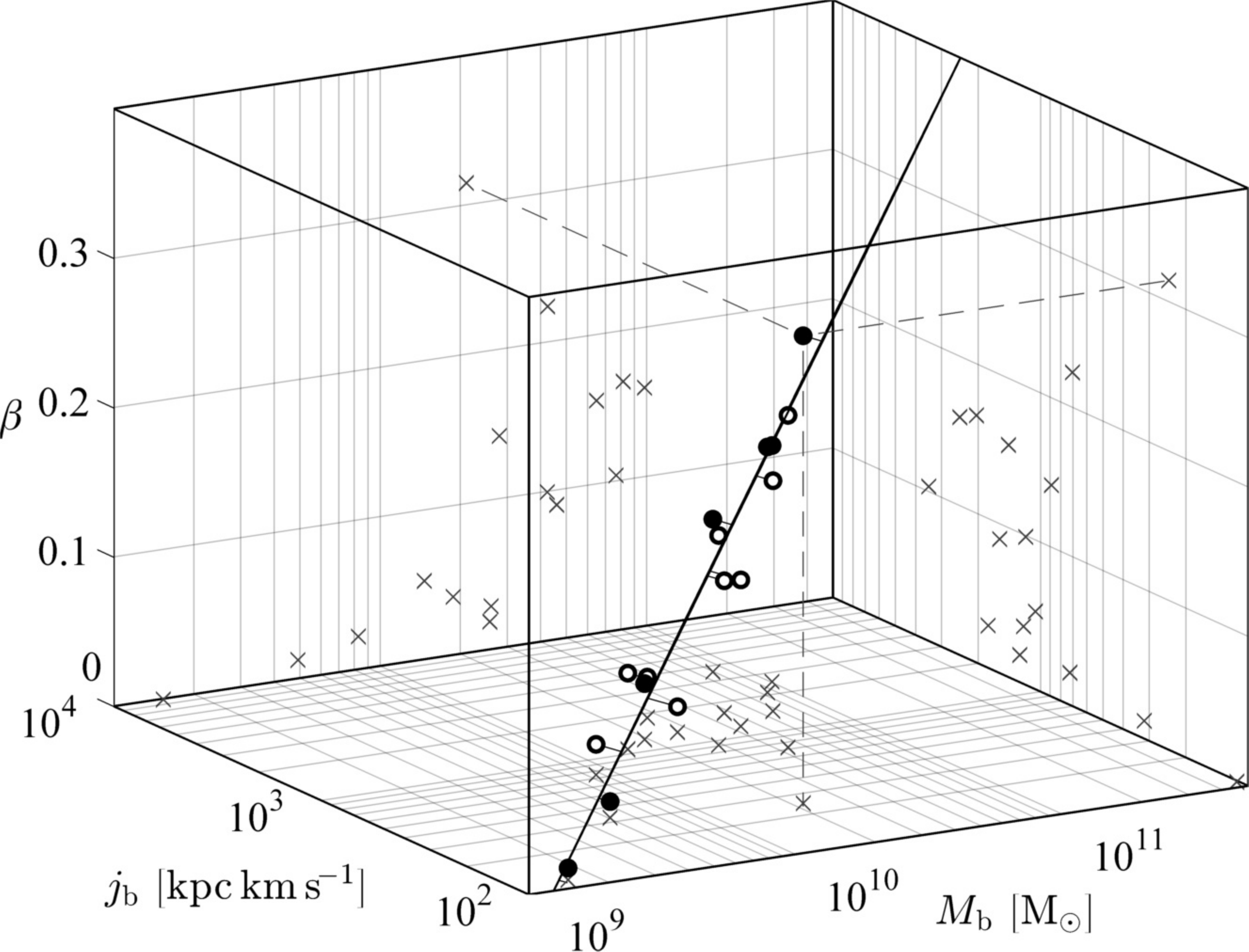}\put(0,71){\normalsize\textbf{(b)}}\end{overpic} \\ [4.0ex]
		\put(10,0){\includegraphics[width=\columnwidth]{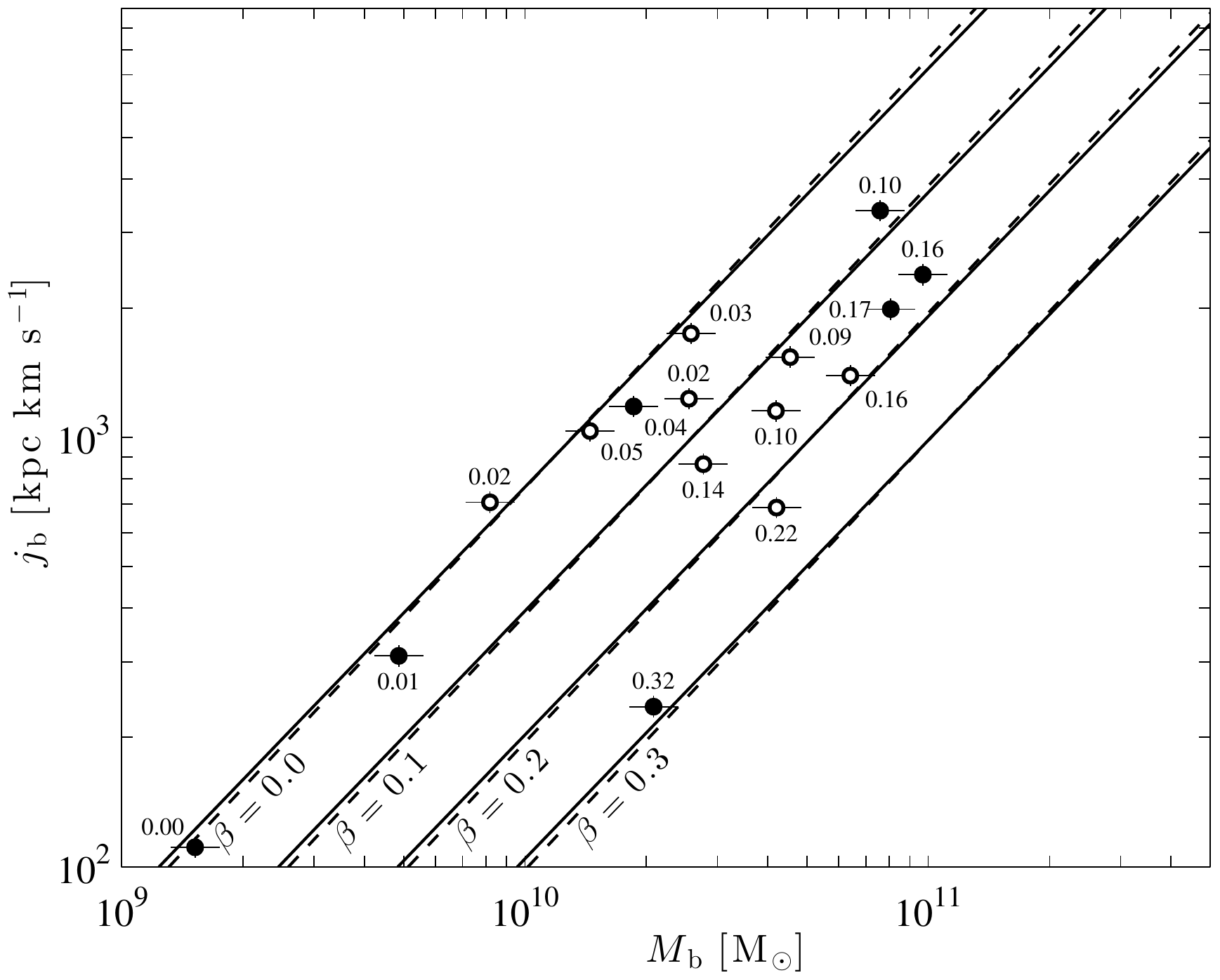}}\put(5,184){\normalsize\textbf{(c)}} &
		\put(10,0){\includegraphics[width=0.99\columnwidth]{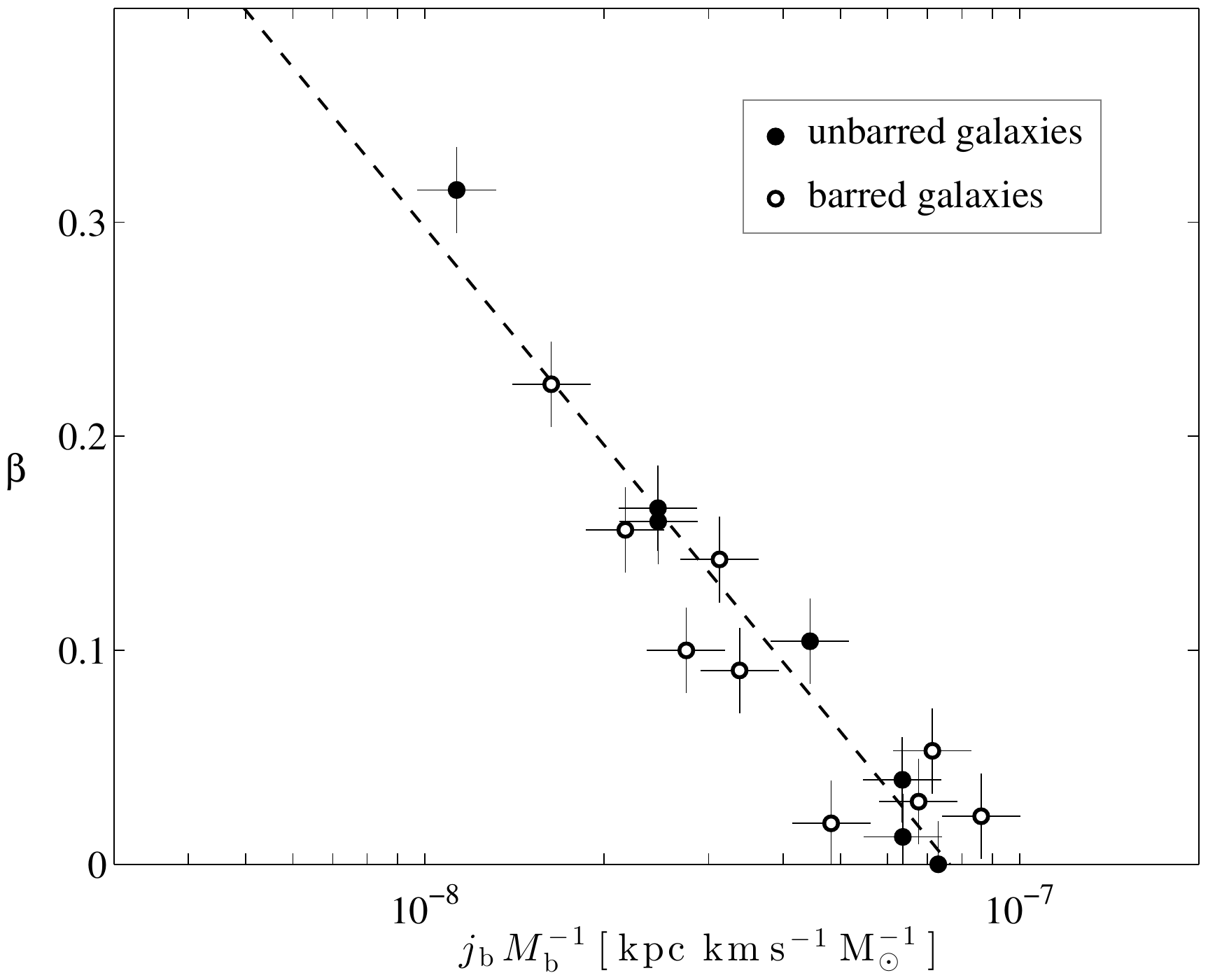}}\put(5,184){\normalsize\textbf{(d)}} \\
	\end{tabular}
	\caption{The four panels show different projections of the 16 spiral galaxies in ($\Mb$,$\jb$,$\bt$)-space. Unbarred and barred galaxies are represented by filled and open circles, respectively. The projections of the 16 data points onto the three planes ($\Mb$,$\jb$), ($\Mb$,$\bt$), and ($\jb$,$\bt$) are represented by the crosses in panel (b). The blue plane in panel (a) is the best trivariate fit to the data (in log-log-lin space), as given in \eq{fit_bt_from_M_and_j}. The same plane is shown edge-on as a solid line in panel (b) and at discrete values of $\bt$ as solid lines in panel (c). The dashed lines, representing \eq{fit_bt_from_jM}, are the best fit to the data when imposing a linear dependence between $\bt$ and $\lg(\jb/\Mb)$, which implies $\jb\propto\Mb$ at any fixed $\bt$. This fit becomes a single dashed line in panel (d) and one dashed line of slope 1 for every value $\bt$ in panel (c). The small numbers next to the data points in panel (c) show their $\bt$ values. Error bars represent standard deviations of the measurement uncertainties. For clarity, these error bars are only displayed in panels (c) and (d). \col}
	\label{fig_MJB}
\end{figure*}

\section{Phenomenology of the $M$-\lowercase{$j$}-$\bt$ relation}\label{section_mj_relation}

This section explores the 3D relationship between $M$, $j$, and the bulge mass fraction $\bt$ of the 16 late-type ($0\leq\bt\lesssim0.3$) galaxies from the THINGS sample considered in this work. Throughout this section, $M$ and $j$ refer to the baryonic $\Mb$ and $\jb$ or the stellar $\Ms$ and $\js$.

\subsection{Fundamental Empirical Relationships}\label{subsection_all_baryons}
Intriguingly, the 16 spiral galaxies turn out to be highly correlated in $(M,j,\bt)$-space. \fig{MJB} shows this space in four alternative projections, revealing that the data form a plane in the space spanned by $\lg\Mb$, $\lg\jb$, and $\bt$, where `lg' denotes the base 10 logarithm. Importantly, this $M$-$j$-$\bt$ relation does not seem to be affected by central bars -- a feature worth investigating in future studies\footnote{Using 10,674 disk galaxies from SDSS, \cite{CervantesSodi2013} found evidence for a dependence of bars on the galactic spin parameter, with unbarred galaxies occupying an intermediate range of spin parameters between short- and long-barred ones.}. The plane can be expressed as
\be\label{eq_fit_bt_from_M_and_j}
	\bt=k_1\lg\left[\frac{M}{10^{10}\msun}\right]+k_2\lg\left[\frac{j}{\rm10^3\unitj}\right]+k_3,
\ee
where $k_1$, $k_2$, and $k_3$ are free parameters, fitted to the data using a trivariate regression (Appendix \ref{appendix_regressions}) that accounts for normal measurement errors in all three dimensions. The best fits are $(k_1,k_2,k_3)=(0.34\pm0.03,-0.35\pm0.04,-0.04\pm0.02)$ if $(M,j)=(\Mb,\jb)$, and $(k_1,k_2,k_3)=(0.31\pm0.03,-0.33\pm0.05,-0.02\pm0.02)$ if $(M,j)=(\Ms,\js)$. The intervals denote 68\% confidence intervals of the correlated uncertainties. \eq{fit_bt_from_M_and_j} is represented by the plane in \figs{MJB}{a}{b}. The correlation between the measured values of $\bt$ and those predicted by \eq{fit_bt_from_M_and_j} is surprisingly high, with a Pearson correlation coefficient of $0.95\!\!$. The reduced $\chi^2$ of the fit is $0.9\!\!$; thus the deviations of the data from the fit are entirely accounted for by measurement errors. In other words, the data is consistent with zero intrinsic scatter off \eq{fit_bt_from_M_and_j}. Another interesting feature is that \eq{fit_bt_from_M_and_j} is irreducible in the sense that it cannot be explained based on the 2D relations $M$-$j$, $M$-$\bt$, and $j$-$\bt$. This is best seen when projecting the data onto the three planes (crosses in \fig{MJB}(b)). In any of these planes, the reduced $\chi^2$ ($14.4$, $11.5$, and $18.1$\!\!) of a linear regression is significantly higher than in 3D.

When discussing the data in the $(M,j)$-plane it is convenient to rewrite \eq{fit_bt_from_M_and_j} as
\be\label{eq_fit_j_from_M_and_bt}
	\frac{j}{10^3\unitj}=k~\xi(\bt)\left[\frac{M}{10^{10}\msun}\right]^\alpha,
\ee
where $\xi(\bt)=\exp[-g\bt]$ (obtained when exponentiating \eq{fit_bt_from_M_and_j}) is a bulge-dependent scaling factor equal to unity in the case of a pure disk ($\bt=0$). The best-fitting parameters are $(k,\alpha,g)=(0.77\pm0.07,0.98\pm0.06,6.65\pm1.02)$ if $(M,j)=(\Mb,\jb)$, and $(k,\alpha,g)=(0.89\pm0.11,0.94\pm0.07,7.03\pm1.35)$ if $(M,j)=(\Ms,\js)$. \eq{fit_j_from_M_and_bt} is shown as solid lines in \fig{MJB}(c) for different values of $\bt$. Interestingly, the exponent $\alpha$ is consistent with $\alpha=1$. Upon imposing $\alpha=1$, the best fit to \eq{fit_j_from_M_and_bt} is $(k,g)=(0.76\pm0.05,6.83\pm0.61)$ for all baryons and $(k,g)=(0.91\pm0.09,7.59\pm0.79)$ for stars only. This fit is shown as dashed lines in \fig{MJB}(c). Given $\alpha=1$, \eq{fit_j_from_M_and_bt} can then be rewritten as
\be\label{eq_fit_bt_from_jM}
	\bt=k_1\lg\left[\frac{j M^{-1}}{10^{-7}\unitj\,\msun^{-1}}\right]+k_2
\ee
with $(k_1,k_2)=(-0.34\pm0.03,-0.04\pm0.01)$ for baryons and $(k_1,k_2)=(-0.30\pm0.03,-0.01\pm0.01)$ for stars. \eq{fit_bt_from_jM} is shown as dashed lines in \fig{MJB}(d).

\subsection{Stars versus Baryons}\label{subsection_stars_vs_baryons}

\begin{figure}[t]
 	\centering
	\includegraphics[width=\columnwidth]{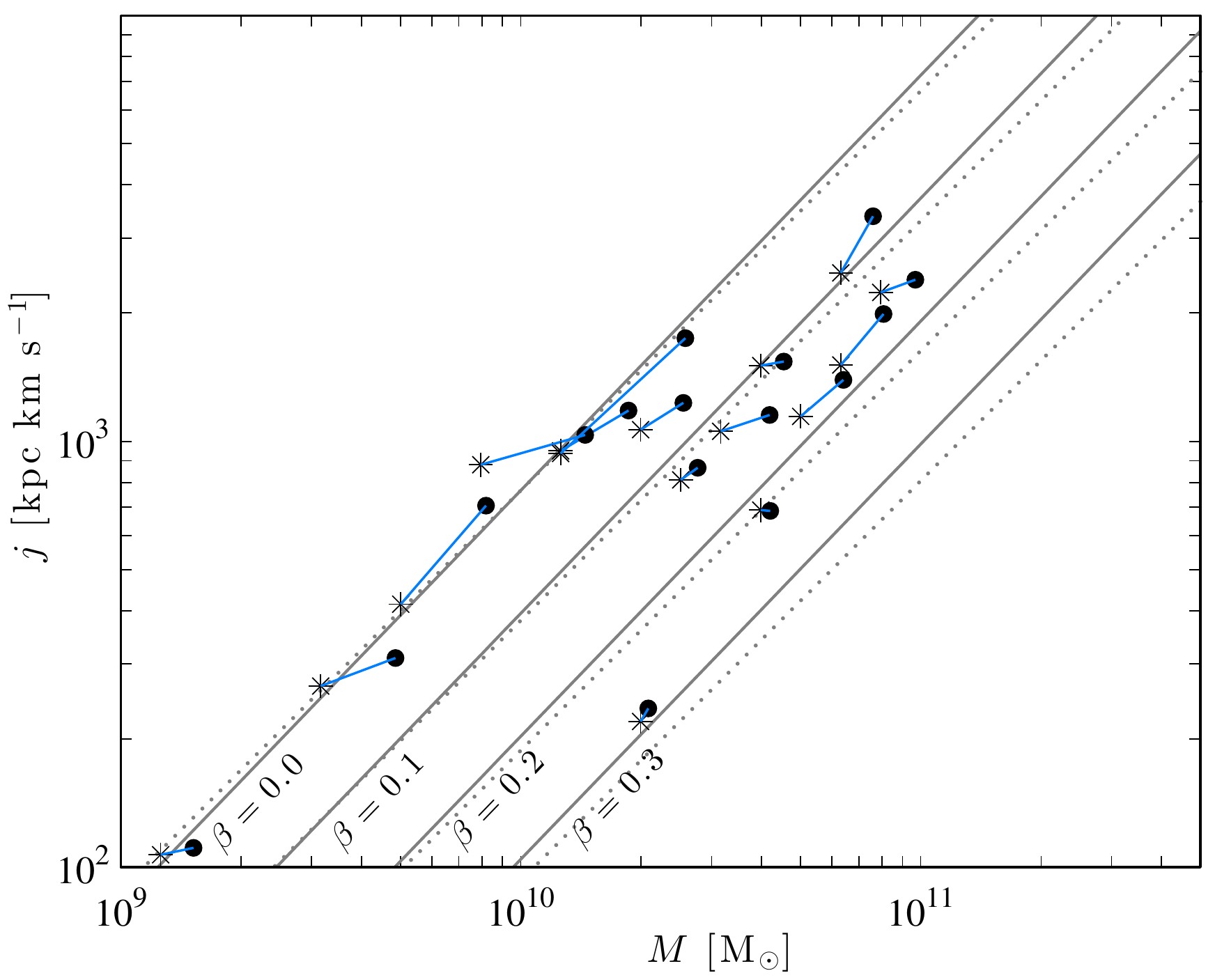}
	\caption{Comparison of the $M$-$j$ distribution for baryons (filled dots) and stars (stars). Identical galaxies are connected by blue lines. Gray lines are the fits (\eq{fit_j_from_M_and_bt}) at different bulge mass fractions $\bt$ for baryons (solid lines) and stars (dotted lines). \col}
	\label{fig_stars_vs_baryons}
\end{figure}

\begin{figure*}[t]
	\centering
	\begin{tabular}{cc}
		\begin{overpic}[width=\columnwidth]{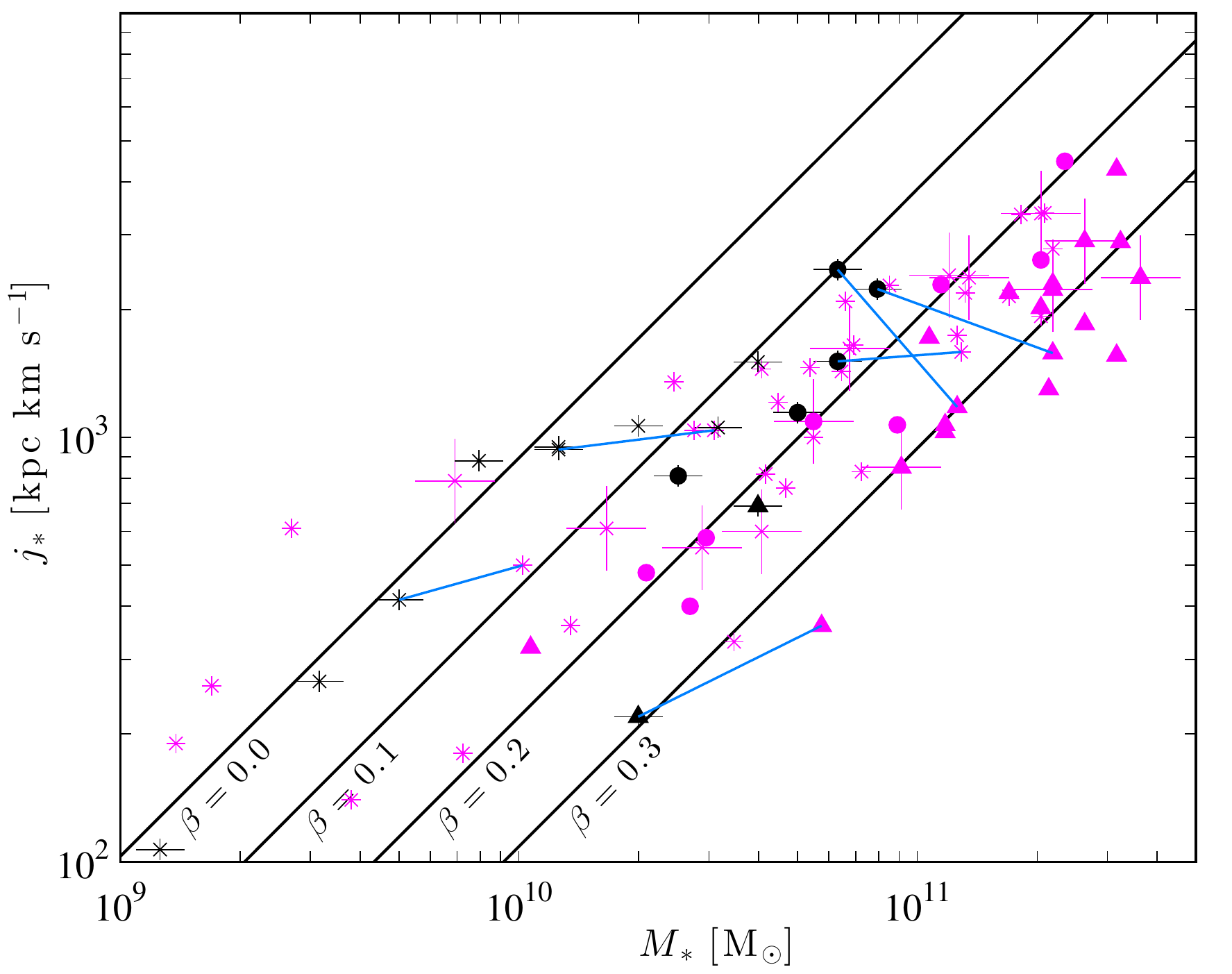}\put(0,74.5){\normalsize\textbf{(a)}}\end{overpic} & 
		\begin{overpic}[width=0.99\columnwidth]{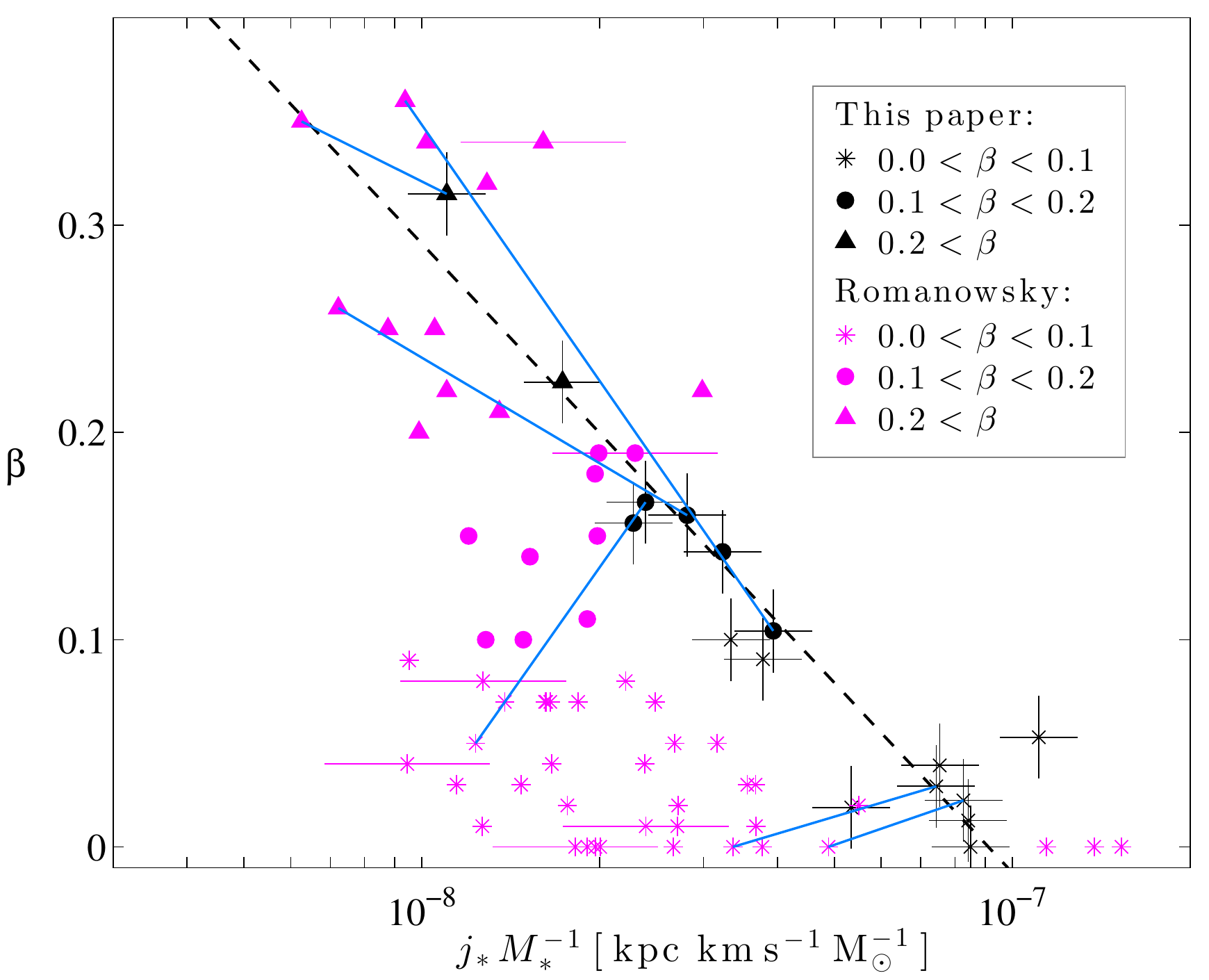}\put(0,74.5){\normalsize\textbf{(b)}}\end{overpic} \\ [4.0ex]
		\put(4,0){\includegraphics[width=\columnwidth]{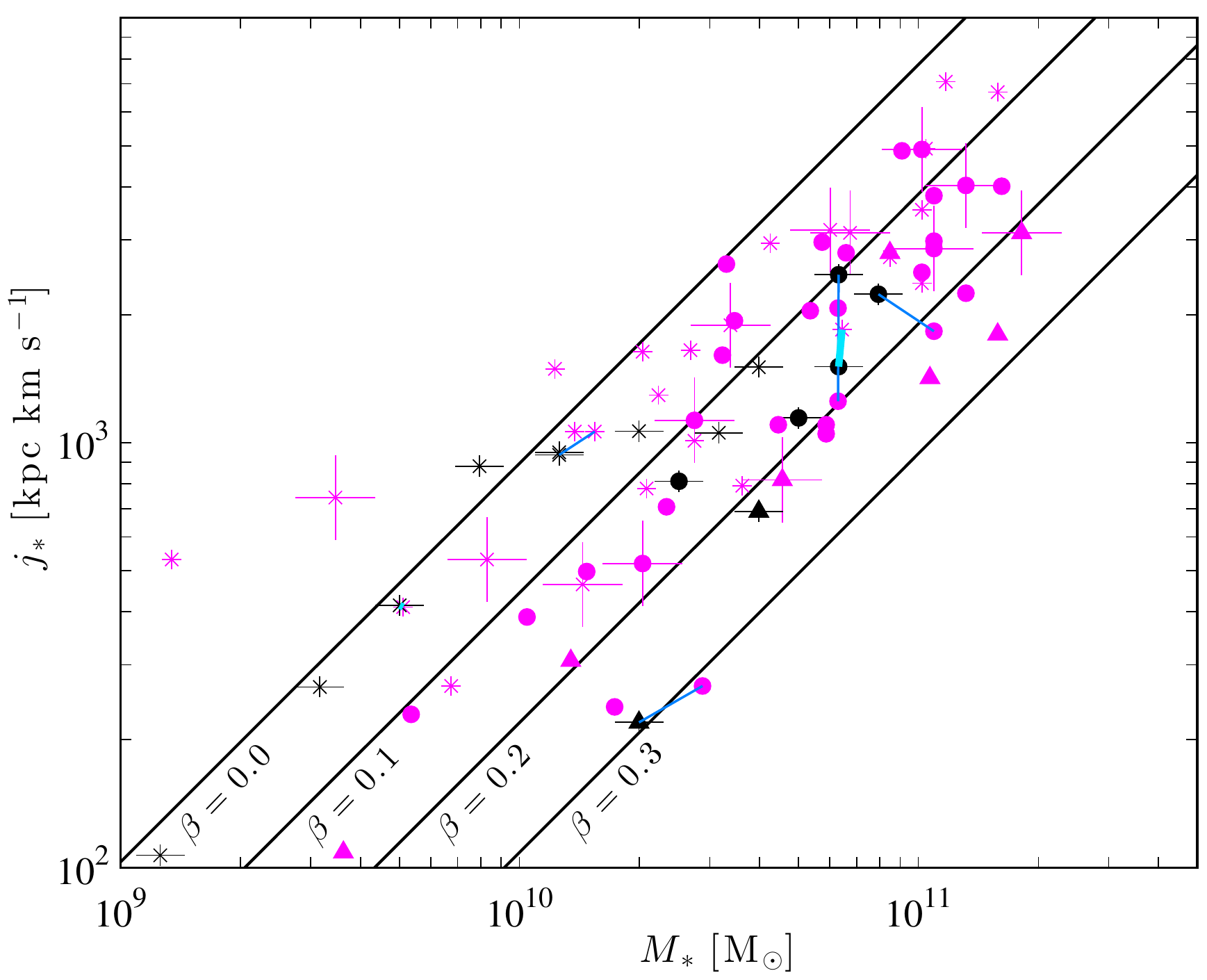}}\put(5,184){\normalsize\textbf{(c)}} &
		\put(4,0){\includegraphics[width=\columnwidth]{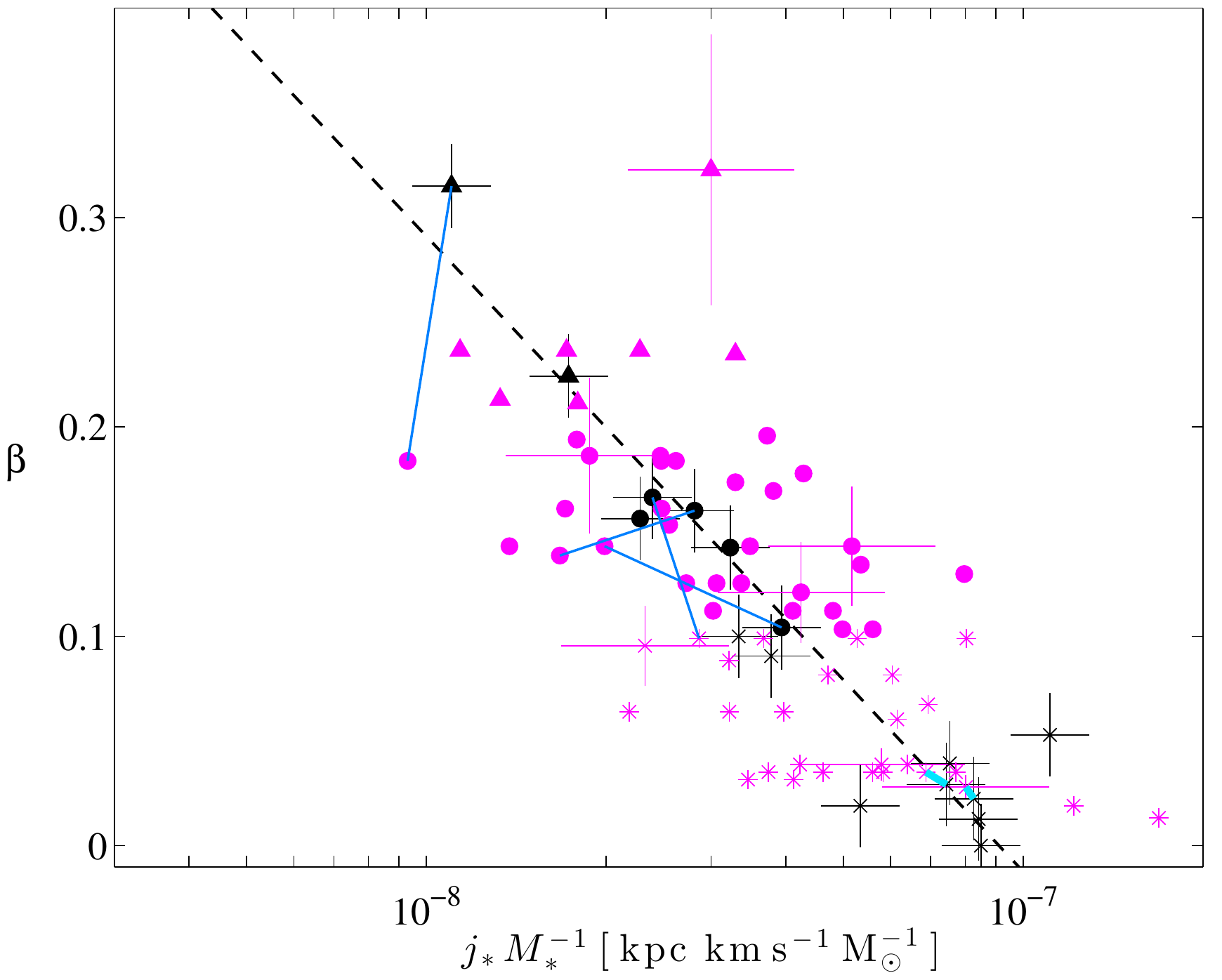}}\put(5,184){\normalsize\textbf{(d)}} \\
	\end{tabular}
	\caption{Comparison of the 16 spiral galaxies in the THINGS sample (black) against the 67 spiral galaxies in the sample of \cite{Romanowsky2012} (pink). The six galaxies present in both samples are connected with blue lines. Left and right panels show two different projections of the ($\Ms$,$\js$,$\bt$)-space. Different symbols separate three ranges of the bulge fraction $\bt$. The THINGS points are identical in the top and bottom panels. They represent $\Ms$ as given by \cite{Leroy2008} and $\js$ and $\bt$ determined in this paper. The Romanowsky points in the top panels are those given in Table~4 of \cite{Romanowsky2012}, while those in the bottom panels have been corrected in all three coordinates to allow a fairer comparison; in particular, the $\js$-values have been rescaled using \eq{fit_j_from_japprox} and the bulge fractions $\bt$ have been computed from the Hubble types (details in \S\ref{subsection_comparison_other_data}). The solid lines in panels (a) and (c) represent the fit of \eq{fit_j_from_M_and_bt} for discrete values of $\bt$, while the dashed lines in panels (b) and (d) represent the fit of \eq{fit_bt_from_jM}. Error bars represent standard deviations of the measurement uncertainties. For clarity, only some error bars are shown. Note that the uncertainties of $\bt$ in \cite{Romanowsky2012} are difficult to estimate; therefore no vertical error bars are shown for the Romanowsky data in panel (b).\col}
	\label{fig_BtoT_comparison}
\end{figure*}

The $M$-$j$-$\bt$ relation turns out to be surprisingly similar for all baryons (star+cold gas) and for stars alone (\fig{stars_vs_baryons}). In fact, the fitting parameters for baryons and stars (given below \eq{fit_bt_from_M_and_j}) are consistent within their uncertainties. This is due to the fact that adding cold gas approximately moves the galaxies in the $(M,j)$-plane along lines of constant $\bt$ (blue lines in \fig{stars_vs_baryons}). In other words, the transition from stars to baryons essentially moves the galaxies inside a fixed $M$-$j$-$\bt$ plane. Formally, the close similarity between the relations $\Mb$-$\jb$-$\bt$ and $\Ms$-$\js$-$\bt$ is due to the fact that $\bt$ varies approximately as $\jb\Mb^{-1}$ coupled with the fact that $\jb\Mb^{-1}\approx\js\Ms^{-1}$. The latter equation is possible, since the contribution of cold gas to the baryon angular momentum $\Jb$ (about 34\% on average) is higher than the contribution of cold gas to $\Mb$ (about 23\%).

The similarity between the relations $\Mb$-$\jb$-$\bt$ and $\Ms$-$\js$-$\bt$ might break down if dwarf galaxies of much higher gas fractions were included. The precise relationship between angular momentum in stars and different cold gas phases will be discussed in a sequel paper.

\subsection{Comparison Against Earlier Studies}\label{subsection_comparison_other_data}

Having established the $M$-$j$-$\bt$ relation in \S\ref{subsection_all_baryons}, we now compare this relation against published data. No significant sample of spiral galaxies with detailed measurements of angular momentum, based on summation of sub-kpc maps, has yet been published. All approximate measurements are restricted to stellar angular momentum without including gas. Thus, this comparison is restricted to samples of approximate stellar angular momentum. The largest and broadest sample was recently published by \cite{Romanowsky2012}, who estimated the stellar angular momenta in a broad mass-range of spiral, lenticular, and elliptical galaxies. Here, we focus on the 67 spiral galaxies listed in Table~4 of \citeauthor{Romanowsky2012}. This table contains bulge mass fractions $\bt$, based on the $r$-band bulge-disk decomposition of \cite{Kent1986,Kent1987,Kent1988}, stellar masses $\Ms$ based on 2MASS $K$-band photometry, and approximate stellar angular momenta $\js$ estimated from global size and velocity measurements (see \S\ref{subsection_approximations}). Given the irreducible 3D-correlation between $\Ms$, $\js$, and $\bt$, the average $\Ms$-$\js$ relation is a poor and potentially misleading estimator for the comparison of two datasets with different $\bt$-distributions. Therefore, the comparison of the 16 THINGS galaxies against the 67 Romanowsky galaxies must be performed in $(\Ms,\js,\bt)$-space or several projections thereof. 

The top panels in \fig{BtoT_comparison} show two projections of the $\Ms$-$\js$-$\bt$ relation: the $\Ms$-$\js$ relation and the $(\js\Ms^{-1})$-$\bt$ relation. These are the same projections as those in the bottom panels of \fig{MJB} (but for stars instead of all baryons). Clearly, the Romanowsky data deviate significantly and systematically from the THINGS data in all three coordinates. This deviation is dominated by differences in measurement techniques, not by systematic differences between the two samples, as can be seen from the six galaxies that are in both samples, connected by lines in \fig{BtoT_comparison}. Upon careful inspection, the following features explain the offset of the Romanowsky points.

\textit{$\Ms$-axis:} Since \citeauthor{Romanowsky2012} use a $K$-band mass-to-light ratio of $1\,\msun/{\rm L}_{\odot,K}$, while THINGS data \citep{Leroy2008} assumed $0.5\,\msun/{\rm L}_{\odot,K}$, the stellar masses of Romanowsky masses must be rescaled by a factor 0.5 for the purpose of this comparison.

\textit{$\js$-axis:} As explained in \S\ref{subsection_approximations}, the approximate values $\tilde\js$ of Romanowsky systematically differ from the fully measured $\js$. This systematic offset can be corrected by rescaling the $\tilde\js$ values using \eq{fit_j_from_japprox}.

\textit{$\bt$-axis:} The bulge mass fractions of the Romanowsky data, which were adopted from \cite{Kent1986,Kent1987,Kent1988}, differ significantly from those of the THINGS data, as revealed by the six overlapping objects in \fig{BtoT_comparison}(b). Their numerical values are:
\begin{equation*}
	\begin{array}{lcccccc}
		\text{NGC:} & 2403 & 2841 & 3198 & 4736 & 5055 & 7331\\
		\bt_{\rm Kent}= & 0.00 & 0.36 & 0.00 & 0.35 & 0.05 & 0.26\\
		\bt_{\rm THINGS}= & 0.02 & 0.10 & 0.03 & 0.32 & 0.17 & 0.16
	\end{array}
\end{equation*}
Kent decomposed the galaxies by performing 2D fits to $r$-band images. They only assumed that the disk and the bulge have elliptical isophotes in projection, without imposing a disk/bulge model. By contrast, most other studies, including this paper, fit a specific disk and bulge model. The comparison of these two methods is difficult, even more so when applied to different wavebands. However, the Kent decompositions can be verified against the more recent 2D bulge-disk decompositions in $H$-band by \cite{Weinzirl2009}. They assumed exponential disks and S{\'e}rsic bulges, analogous to our decomposition of the THINGS galaxies. Of their 143 galaxies, five overlap with those in the Romanowsky sample\footnote{None of the 143 galaxies in the sample of \cite{Weinzirl2009} overlaps with the 16 THINGS galaxies used here.}. The respective bulge mass fractions disagree considerably:
\begin{equation*}
	\begin{array}{lccccc}
		\text{NGC:} & 1087 & 2775 & 4062 & 4698 & 7217\\
		\bt_{\rm Kent}= & 0.00 & 0.20 & 0.03 & 0.55 & 0.25\\
		\bt_{\rm Weinzirl}= & 0.00 & 0.61 & 0.02 & 0.22 & 0.54
	\end{array}
\end{equation*}
Due to this discrepancy, the values $\bt_{\rm Kent}$ adopted in the Romanowsky sample are not appropriate for the purpose of comparison with the present data. We therefore re-estimate the bulge mass fractions of the Romanowsky galaxies from their numerical Hubble types $T$ (drawn from HyperLeda \citep{Paturel2003} and listed in Table 4 of \cite{Romanowsky2012}). To compute $\bt$ from $T$, we use the mean $T$-$\bt$ relation of \cite{Weinzirl2009}, shown as blue squares in their Figure 14\footnote{For fractional values of $T$, the values of $\bt$ are interpolated linearly between the neighboring integers of $T$. For $T\geq7$ (Sd-Sm), the observed trend for $0\leq T<7$ is extrapolated using the fit $\bt=[(10-T)/16]^{2.5}$, but this extrapolation has little bearing as it only concerns galaxies with $\bt<0.015$.}.

The three adjustments of $\Ms$, $\js$, and $\bt$ in the Romanowsky data are justified and necessary for a fair comparison with the present study. Given these adjustments, the data become consistent with the THINGS data (\fig{BtoT_comparison}, lower panels). In fact, the trivariate fit of \eq{fit_j_from_M_and_bt} to the Romanowsky data with assumed statistical uncertainties of 0.1 dex in $\Ms$ and $\js$ and 20\% for $\bt$, gives $(k,\alpha,g)=(0.99\pm0.15,0.92\pm0.06,7.63\pm0.99)$ in full agreement with the respective parameters of the THINGS galaxies for $(M,j)=(\Ms,\js)$. The reduced $\chi^2$ of this fit is $ 1.7$; thus the scatter of the Romanowsky data is roughly accounted for by observational uncertainties. In this sense the Romanowsky data fully supports the scaling relations of \S\ref{subsection_all_baryons}.


\section{Discussion of the 2D $M$-\lowercase{$j$} relation}\label{section_discussion2D}

In preparation for discussing the full $M$-$j$-$\bt$ relation (Section \ref{section_discussion3D}), this section discusses the distribution of the spiral galaxies in the ($M$,$j$)-plane, relative to predictions from a simplistic analytical model.

\subsection{The ($M$,$j$)-plane in Basic CDM}\label{subsection_interpretation_cdm}

In the model of a singular isothermal spherical CDM halo \citep{Mo1998} of truncation radius $\Rh$ and dynamical mass $\Mh$, Newtonian gravity sets the circular velocity to
\be\label{eq_Vh_isothermal}
	\Vh=(G\Mh/\Rh)^{1/2},
\ee
where $G$ denotes the gravitational constant. For $\Vh$ to be constant (isothermicity), the mass density needs to vary as $\rho(r)=\Vh^2(4\pi G r^2)^{-1}$ $\forall r\leq\Rh$. Thus the potential energy becomes $E_{\rm pot}=-\Mh\Vh^2$. Following the virial theorem ($2E_{\rm kin}=-E_{\rm pot}$), the total energy is
\be\label{eq_Eh_isothermal}
	\Eh = -0.5\Mh\Vh^2.
\ee
Halos are embedded in the cosmic background field of mean density $\rho_c=3H^2(8\pi G)^{-1}$, where $H$ is the Hubble `constant' at the considered epoche. The halo radius $\Rh$ can then be defined as the radius to which orbits are approximately virialized. In the spherical collapse model \citep{Cole1996}, the mean density enclosed by $\Rh$ is about $200\rho_c$, thus $\rho(\Rh)=(200/3)\rho_c$. It follows that
\be\label{eq_Rh_isothermal}
	\Rh^3=10^{-2}GH^{-2}\Mh.
\ee
Equations (\ref{eq_Vh_isothermal}), (\ref{eq_Eh_isothermal}), and (\ref{eq_Rh_isothermal}) are the essential scaling relations of the isothermal CDM halo. This model is manifestly scale-free (at fixed $H$) in that all global quantities depend on a single scale-factor, e.g., on $\Rh$, via
\be\label{eq_fundamental_scalings}
	\Vh\propto\Rh\propto\Mh^{1/3}\propto|\Eh|^{1/5}.
\ee

When dealing with the halo angular momentum $\Jh$, the spin parameter \citep{Steinmetz1995} 
\be\label{eq_spin_parameter}
	\lambda\equiv\Jh|\Eh|^{1/2}G^{-1}\Mh^{-5/2}
\ee
has the advantage of being approximately invariant during the growth of a halo in the absence of major mergers (Fig.~1 in \citealp{Stewart2013}). Combining \eq{spin_parameter} with the scaling equations of the isothermal halo,
\be\label{eq_j_theory}
	\jh\equiv \frac{\Jh}{\Mh} = \sqrt{2}\,\lambda\,\Rh\Vh = \frac{\sqrt{2}\,\lambda\,G^{2/3}}{(10H)^{1/3}}\,\Mh^{2/3}.
\ee
If the baryon angular momentum remains conserved during galaxy formation, then the initial equality $\jb=\jh$ for a uniform mixing of baryons and dark matter applies at all times. More generally, we can define the ratio $\fj\equiv\jb/\jh$, which is unity in the conserved case. Further introducing the baryon mass fraction $\fb\equiv\Mb/\Mh$,
\be\label{eq_jb_theory}
	\jb = \frac{\sqrt{2}\,\lambda\,\fj\,\fb^{-2/3}\,G^{2/3}}{(10H)^{1/3}}\,\Mb^{2/3}.
\ee
Adopting the local $H=70~\kms~\rm Mpc^{-1}$ and conventional units, \eq{jb_theory} becomes
\be\label{eq_jb_theory_units}
	\frac{\jb}{10^3\,\unitj} = 1.96\lambda\fj\fb^{-2/3}\left[\frac{\Mb}{10^{10}\,\msun}\right]^{2/3}.
\ee
This equation is equivalent to Equation (15) of \cite{Romanowsky2012} upon adopting the same $H$ and substituting $\fb=f_{\rm b} f_\star$, where $f_{\rm b}=0.17$ is the universal baryon fraction \citep{Komatsu2011}.

\begin{figure}[t]
	\includegraphics[width=\columnwidth]{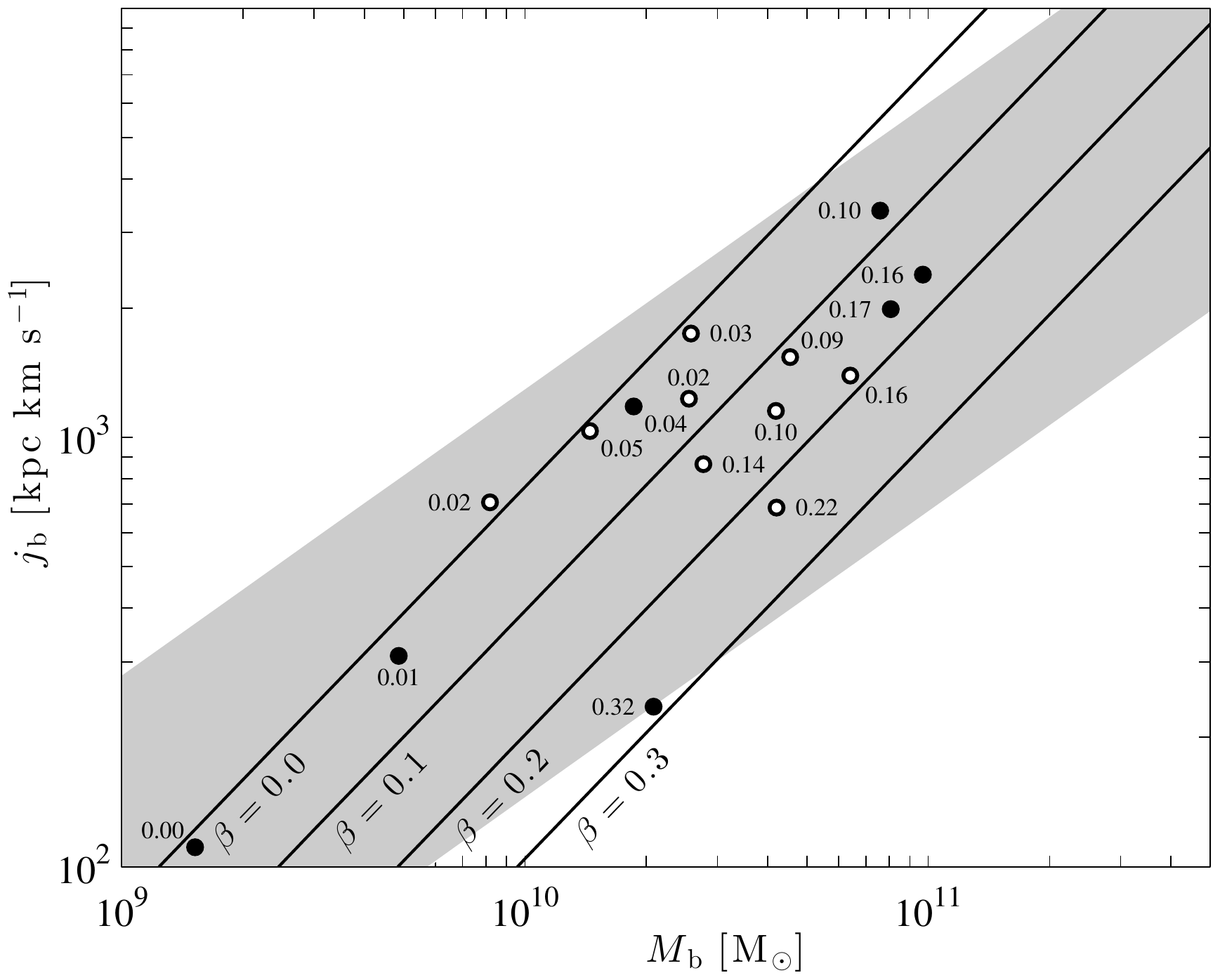}
	\caption{CDM predicts that regular galaxies in isolated halos fall inside the shaded region of the $(\Mb,\jb)$-plane, given by \eq{jb_theory_units}. This region has an average slope of $\alpha=2/3$. The data agrees with this prediction, although for a fixed bulge fraction $\bt$ the power-law index is stepper ($\alpha\approx1$, solid lines). Points and lines are the same as in \fig{MJB}(c), which shows the errors bars.}
	\label{fig_jm_diagram_CDM}
\end{figure}

To compare \eq{jb_theory_units} against the THINGS data, the dimensionless parameters need to be given sensible values. The spin parameter $\lambda$ can be determined from cosmological simulations that tackle the formation of halos, including the tidal build-up of angular momentum. $N$-body simulations find present-day values around $\lambda\approx0.04$ with an intrinsic scatter of about 0.02 and no significant correlation to $\Mh$ \citep{Maccio2008,Knebe2008}. The baryon fraction $\fb$ depends on the galaxy mass and is maximal for intermediate, Milky Way mass galaxies \citep{McGaugh2010,Behroozi2013}. The mean of the stellar mass considered here being approximately equal to that of the Milky Way, we adopt the constant\footnote{A variable value $\fb(M)$, fitted to available data slightly bends the gray-shaded zone of \fig{jm_diagram_CDM} without changing the conclusions.} value of the Milky Way, estimated\footnote{Based on the empirical values of $\Mh$ \citep{McMillan2011}, $\Ms$ \citep{Flynn2006}, $\Mha$ (fit to $\Sha(r)$ in \citealp{Kalberla2008}), and $\Mhm$ (fit to $\Shm(r)$ in Table~3 of \citealp{Sanders1984}). Explicit values given in Table~1 of \cite{Obreschkow2011c}.} to $\fb\approx0.05$. Regarding the spin fraction $\jb$, high-resolution simulations of four Milky Way type galaxies \citep{Stewart2013} find present-day values of $\fj\approx1$ within about 50\%. Given those choices, \smash{$1.96\lambda\fj\fb^{-2/3}$} can vary between $0.14$ and $1.3$, spanning the gray-shaded zone of \fig{jm_diagram_CDM}.

In summary, isolated spiral galaxies, evolved without major mergers, abnormal feedback, or otherwise exotic histories, are predicted to lie in the shaded zone of \fig{jm_diagram_CDM}. This prediction is consistent with the data. Coupling this prediction of a mean relation $\jb\propto\Mb^\alpha$, where $\alpha=2/3$, with the empirical finding of $\alpha\approx1$ for fixed $\bt$'s (solid lines in \fig{jm_diagram_CDM}), implies that more massive spiral galaxies tend to have higher bulge fractions than less massive ones. This trend qualitatively agrees with observations of the stellar mass function split into Sa, Sb, Sc, and Sd types \citep{Read2005}.

\subsection{Linking the ($M$,$j$)-plane to Classical Scaling Laws}

\begin{figure*}[t]
	\centering
	\includegraphics[width=15cm]{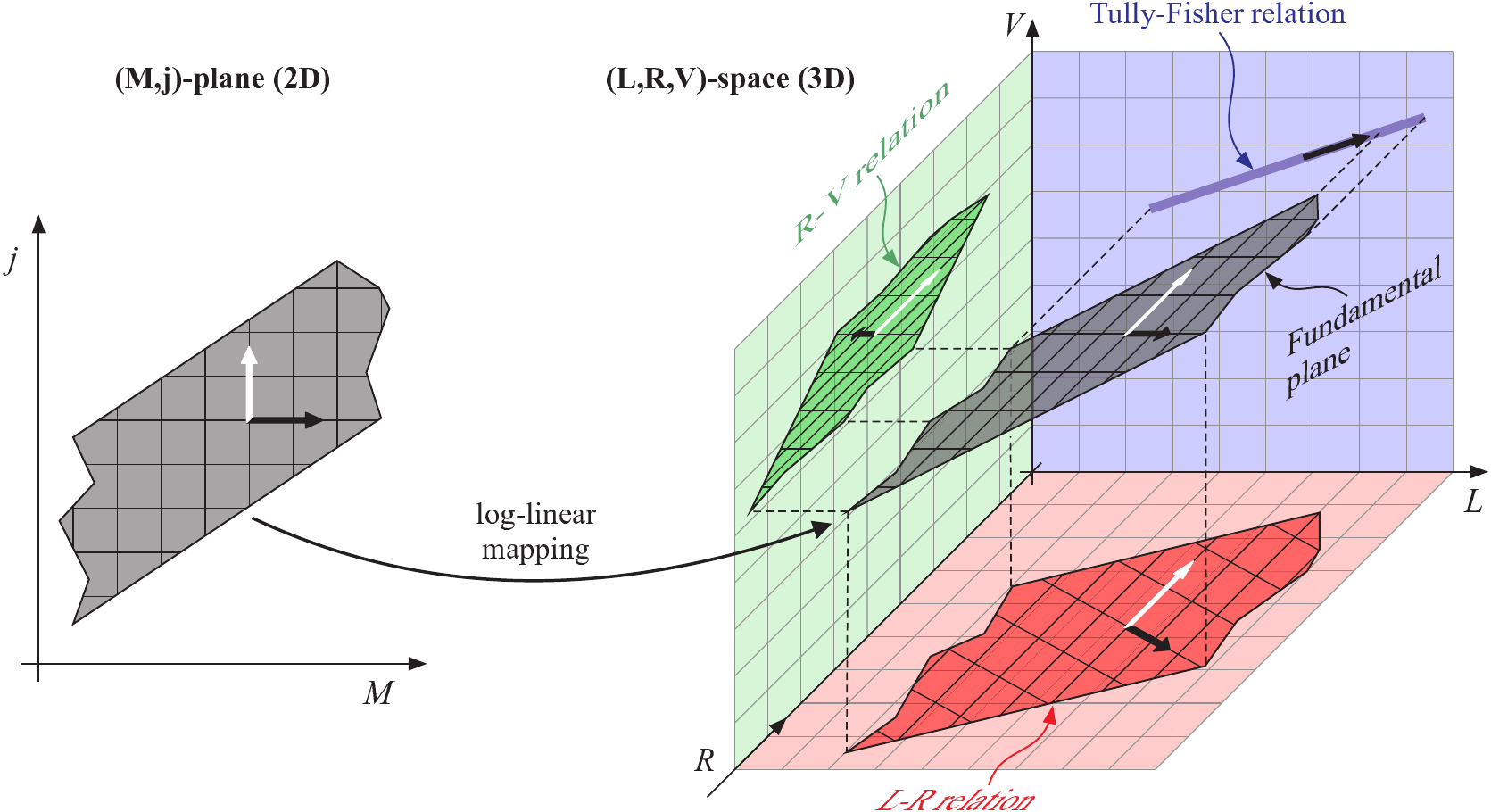}
	\caption{In the model of an exponential disk inside a spherical CDM halo, the FP of spiral galaxies can be understood as a mapping of the ($M$,$j$)-plane (left) into $(L,R,\vflat)$-space (right) via Equations~(\ref{eq_fp_transformations}). Projections of the FP onto the $(L,R)$-plane (red), $(R,V)$-plane (green), and $(V,L)$-plane (red), then gives rise to three classical scaling relations, given in Equations~(\ref{eq_classical_relations}). Of these relations, the $V$-$L$ relation -- the TF relation -- has the smallest scatter, because it is a nearly edge-on projection of the FP.}
	\label{fig_scalings}
\end{figure*}

\begin{figure*}[t]
	\centering
	\includegraphics[width=\textwidth]{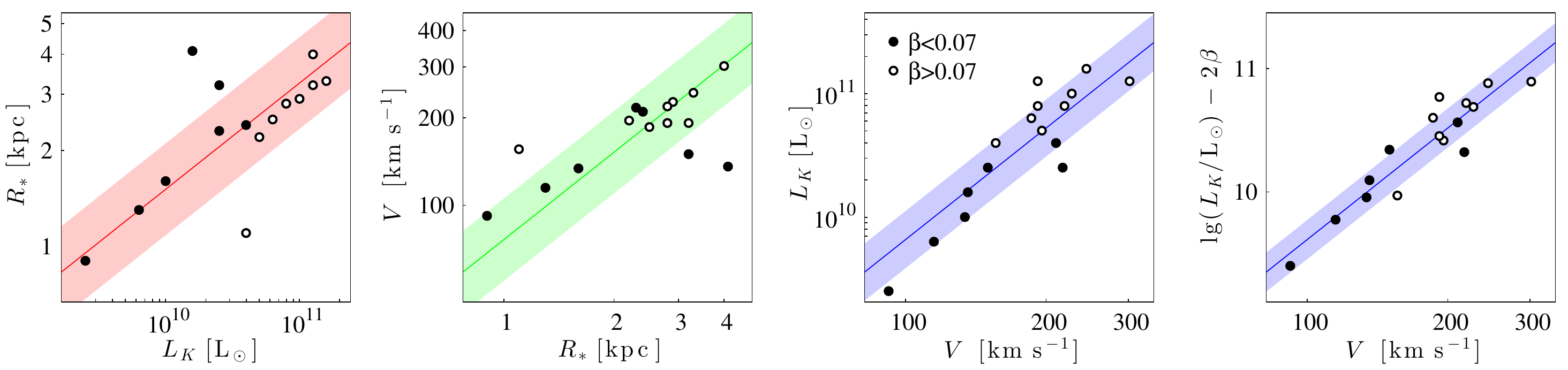}
	\caption{The 16 spiral THINGS galaxies shown in four different planes. The first three panels depict the orthogonal projections of the FP plane, shown in \fig{scalings} with matching colors. The fourth panel is identical to the third (the TF relation), except for an additional $\bt$-term on the vertical scale that factors out the morphology-dependence of the TF relation, thus reducing its scatter. Solid lines are power-laws with zero-points fitted to the data and fixed exponents of $1/3$, $1$, $3$, and $3$, respectively, as predicted by the scale-free model (\eq{classical_relations}). Shaded regions represent 1-$\sigma$ scatter.}
	\label{fig_classical_scalings}
\end{figure*}

The ($M$,$j$)-plane is linked to the fundamental plane (FP) for spiral galaxies \citep{Koda2000,Han2001,Courteau2007}, a 3D relation between total luminosity $L$, disk scale radius $R$, and asymptotic velocity\footnote{Sometimes the integrated velocity dispersion is considered rather than $\vflat$, but in the present context the FP refers to the $L$-$R$-$\vflat$ relation.} $\vflat$, forming a plane in log-space. Projected onto 2D (\fig{scalings}), the FP reduces to the $L$-$R$ relation, the $R$-$\vflat$ relation, and the $\vflat$-$L$ relation. The latter, known as the Tully-Fisher (TF) relation, appears to be a nearly edge-on projection of the FP \citep{Shen2002}.

In the scale-free approximation of Section \ref{subsection_interpretation_cdm}, a direct link between the FP and the ($M$,$j$)-plane appears, since the three quantities $L$, $R$, and $V$ scale with $M$ and $j$ (here, $M\approx\Mb\approx\Ms$, $j\approx\jb\approx\js$, and $R\approx\Rs$). First, the luminosity is a linear proxy of mass, $L\propto M$. Second, in an exponential disk at constant circular velocity, the scale radius becomes $R=j/(2\vflat)$ (\eq{japprox1}). Third, the velocity $\vflat$ can be approximated by the halo velocity $\Vh\propto\Mh^{1/3}$ (\eq{fundamental_scalings}); thus, $\vflat\propto M^{1/3}$ when assuming a constant disk mass fraction $M/\Mh$. Hence the transformation $(M,j)\mapsto(L,R,\vflat)$ writes
\be\label{eq_fp_transformations}
	L\propto M,~~~R\propto j M^{-1/3},~~~\vflat\propto M^{1/3}.
\ee
This mapping is sketched schematically in \fig{scalings}. It implies that one can reconstruct the FP of spiral galaxies from their distribution in the ($M$,$j$)-plane. This distribution is described by $j=kM^{2/3}$ (\eq{jb_theory}), where $k$ is a $\lambda$-dependent, scattered parameter (gray shading in \fig{jm_diagram_CDM}). Combining $j=kM^{2/3}$ with Equations~(\ref{eq_fp_transformations}), the projected relations of the FP become
\be\label{eq_classical_relations}
	R\propto k L^{1/3},~~~\vflat\propto k^{-1}R,~~~L\propto\vflat^3.
\ee

These scalings are remarkably similar to those found by \cite{Courteau2007} in $I$-band for a sample of 1,300 spiral galaxies of all Hubble types (S0a-Sm). Their fits are $R\propto L^{0.32\pm0.02}$ (scatter $\sigma_{\ln R}=0.33$), $R\propto\vflat^{1.10\pm0.12}$ ($\sigma_{\ln R}=0.38$), $\vflat\propto L^{0.29\pm0.01}$ ($\sigma_{\ln\vflat}=0.13$). The scatter of the third scaling -- the TF relation -- is significantly smaller than that of the other two, relative to the range spanned by the data (Figure~3 in \citeauthor{Courteau2007}). This difference in scatter is elegantly explained by the fact that the first two relations in Equations~(\ref{eq_classical_relations}) depend on $k$, while the TF relation does not, as it is an exactly edge-on projection of the FP in our simplistic model.

The three relations of \eq{classical_relations} are consistent with the present sample, where $L=L_K$ and $R=\Rs$, as shown in \fig{classical_scalings} (first three panels). Solid lines are power-laws with zero-points fitted to the data and exponents fixed according to \eq{classical_relations}. Shaded regions denote standard deviations. The location of a galaxy in these planes depends on its position in the ($M$,$j$)-plane, which systematically depends on $\bt$ (Section \ref{section_mj_relation}); thus the visible offset from between open and filled points in \fig{classical_scalings}. The TF relation exhibits the smallest scatter relative to the range of the data. This had to be expected from the TF being an edge-on projection of the FP in the model discussed so far. In reality, the TF relation is not exactly an edge-on projection of the FP, as explained by the more detailed theory of \cite{Shen2002}. Therefore, the offset of galaxies from the mean the TF relation correlates with their location in the ($M$,$j$)-plane, thus with $\bt$ (hence departing from \eq{classical_relations}, right). This explains the slight morphology-dependence of the TF relation \citep{Kannappan2002}, also visible in the present sample. In fact, we can minimize the scatter of the TF relation by heuristically substituting $\lg L_K$ for $\lg L_K-u\bt$ with $u\approx2$ (last panel in \fig{classical_scalings}).

In summary, within the model of an exponential disk inside a CDM halo, \textit{the FP results from mapping the 2D ($M$,$j$)-plane into 3D $(L,R,\vflat)$-space} via Equations~(\ref{eq_fp_transformations}). This mapping approximately explains the three classical scaling relations that are the 2D projections of the FP, such as the TF relation. The morphology dependence of these three relations can then be traced back to the $M$-$j$-$\bt$ relation established empirically in Section \ref{section_mj_relation}.


\section{Discussion of the 3D $M$-\lowercase{$j$}-$\bt$ relation}\label{section_discussion3D}

In the previous section, the $\bt$-dependence of $M$-$j$ relation was considered an empirical fact, useful to explain the morphology dependencies of other relations. Any physical explanation of the full $M$-$j$-$\bt$ relation is expected to answer questions such as: What physical processes dominate this relation? Is it self-regulated such that galaxies offset from the relation will evolve back onto it? Which of the quantities $M$, $j$, and $\bt$ are the cause and the effect? These questions call for a model that can reproduce the $M$-$j$-$\bt$ relation from more fundamental scaling laws, time-independent physics (e.g., conservation laws and stability criteria), or time-dependent models (e.g., semi-analytic models or hydrodynamic simulations). In \S\ref{subsection_independent_models}, an explanation based on independent $M$-$j$ relations for disks and bulges is shown to be at odds with the data. A path toward an alternative explanation is then discussed in \S\ref{subsection_idea_for_explanation}.

\subsection{Failure of the Two-component Model}\label{subsection_independent_models}

When discussing the Hubble type-dependence of the $M$-$j$ relation, \cite{Fall1983} and \cite{Romanowsky2012} invoked the idea that this dependence might result from different, fixed $M$-$j$ relations for pure disks and pure bulges. While this idea might be valid for classical bulges in bulge-dominated systems, the data of this paper dispels the hope for such an elegant explanation in the case of spiral galaxies with smaller (pseudo-)bulges.

\begin{figure}[t]
	\includegraphics[width=\columnwidth]{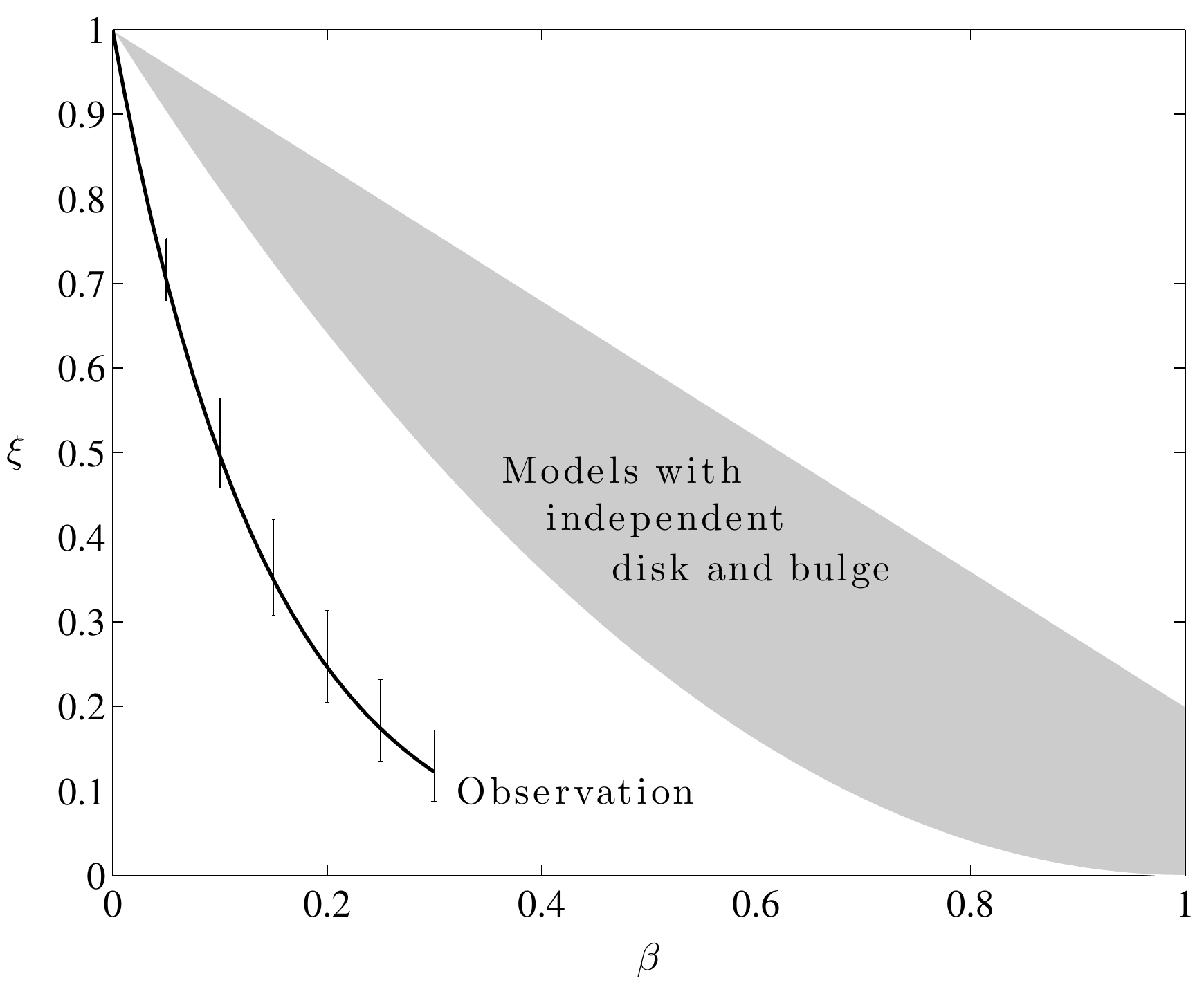}
	\caption{Function $\xi(\bt)$ defined in \eq{jb_models}. The solid line represents the measured function for stars (see \eq{fit_j_from_M_and_bt}), whereas the shaded region represents the plausible range if the $M$-$j$-$\bt$ relation were explainable based on independent, fixed $M$-$j$ relations for pure disks and pure bulges.}
	\label{fig_xi}
\end{figure}

Let us assume -- ad absurdum -- that disks and bulges do indeed obey independent $M$-$j$ relations. This assumption can be understood in two ways, formalized via the following models. In `model 1', disk and bulge are strictly independent in the sense that they obey different relations $\jdisk=k\Mdisk^\alpha$ and $\jbulge=fk\Mbulge^\alpha$ with constants $k>0$ and $f>0$. In `model 2', the angular momenta of disk and bulge both depend on the same total mass $M=\Mdisk+\Mbulge$, i.e., $\jdisk=k M^\alpha$ and $\jbulge=fk M^\alpha$. In both models, the total specific angular momentum $j=(1-\bt)\jdisk+\bt\jbulge$ becomes
\be\label{eq_jb_models}
	j = k~\xi(\bt)\,M^\alpha,
\ee
where $\xi(\bt)=(1-\bt)^{1+x}+f\bt^{1+x}$ with $x=\alpha$ for model 1 and $x=0$ for model 2. Intermediate models can then be obtained by choosing $0<x<\alpha$. Considering this range for $x$, choosing $f$ between $f=0$ (zero-rotation bulge model of \citealp{Mo1998}) and $f=0.2$ \citep[empirical value of][]{Fall2013}, and adopting the empirical $\alpha\approx1$ (\eq{fit_j_from_M_and_bt}), implies that $\xi(\bt)$ falls within the shaded region of \fig{xi}. By contrast, $\xi(\bt)=\exp[-g\bt]$ determined empirically (see below \eq{fit_j_from_M_and_bt}) varies as the solid line in \fig{xi}. This measurement is clearly inconsistent with any plausible model of independent disk and bulge relations -- an unrealistic $\alpha\approx6$ or $f<0$ would be required to match up the model with the data. Therefore, the initial assumption of independent $M$-$j$ relations for disks and bulges cannot be true.

\begin{figure}[t]
	\includegraphics[width=\columnwidth]{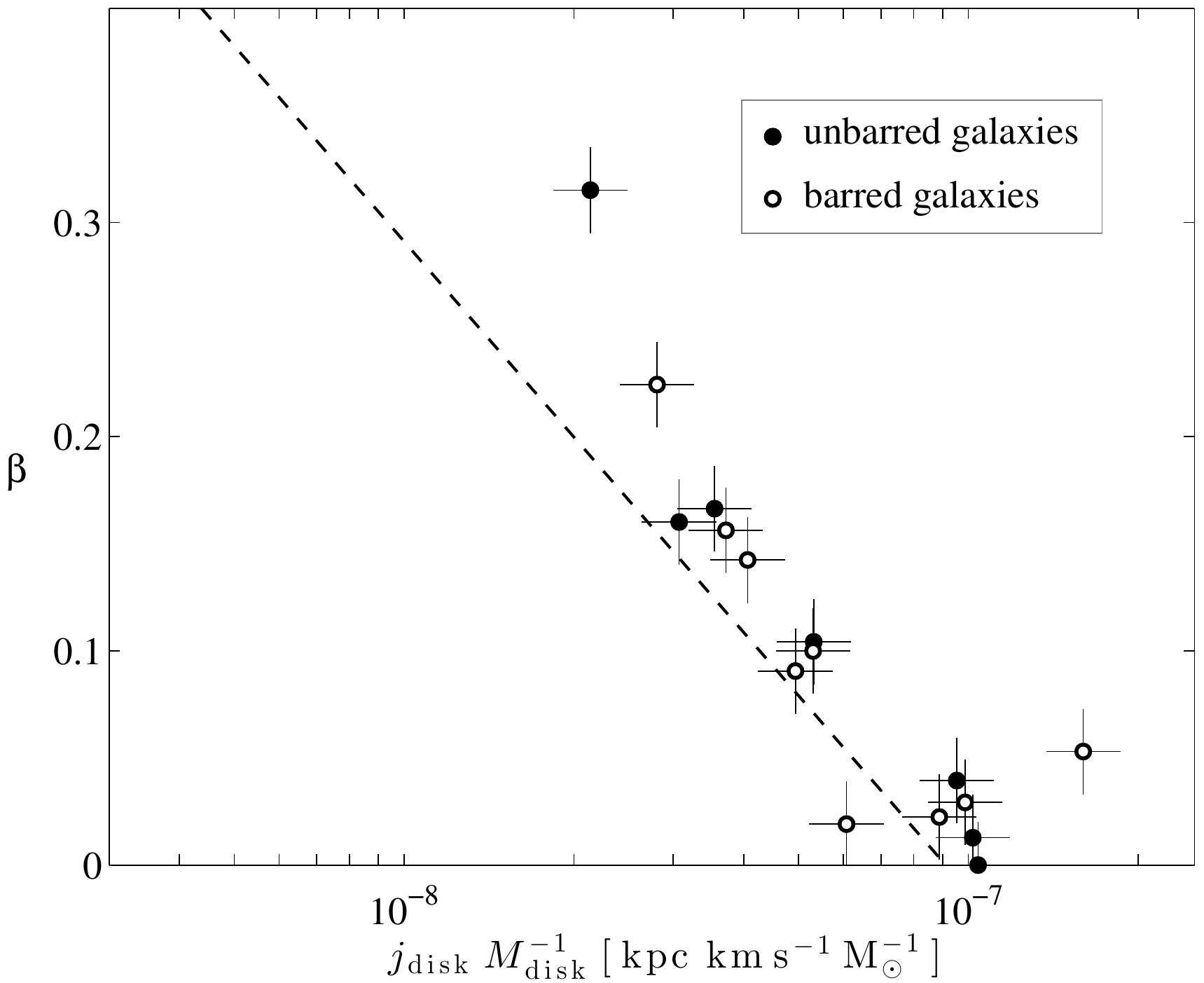}
	\caption{Projection of the $M$-$j$-$\bt$ relation using only the mass and angular momentum of the stellar disk, excluding the bulge. The dashed line shows the best fit to the data for the whole bulge-disk system; this is the same dashed line as in \fig{BtoT_comparison}(d). This relation is offset to the left, because the small $j$ of the bulge reduces the value of $j M^{-1}$ compared to $\jdisk\Mdisk^{-1}$.}
	\label{fig_mjb_disk}
\end{figure}

This conclusion can be confirmed explicitly by measuring the $\Ms$-$\js$-$\bt$ relation of the disk component only. The stellar mass of the disk $\Mdisk=\bt\Ms$ is drawn directly from the stellar bulge-disk decompositions (Appendix \ref{appendix_disk_bulge}). The specific stellar angular momentum of the disk $\jdisk$ is computed via \eq{j}, substituting $\Sigma(r)$ for the disk stellar mass surface density, again drawn from our bulge-disk decompositions. \fig{mjb_disk} shows the resulting relation projected onto the $(\jdisk\Mdisk^{-1},\bt)$-plane. It turns out that \smash{$\jdisk\Mdisk^{-1}$} correlates strongly with $\bt$, hence explicitly rejecting the model of a fixed $M$-$j$ relation for the disk component. Disks with more massive bulges in their centers have lower specific angular momentum, thus smaller radii for a given mass.

In summary, disks `know' about the bulges via their angular momentum -- an interesting feature that must be accounted for by any model of the $M$-$j$-$\bt$ relation.

\subsection{Surface Density Approach to the $M$-$j$-$\bt$ Relation}\label{subsection_idea_for_explanation}

Late-type galaxies grow their (pseudo-)bulges \textit{in situ} \citep{Elmegreen2008,Weinzirl2009}, rather than via major mergers (mass ratios $>0.3$) thought to produce the classical bulges of early-type galaxies \citep{Koda2009}. Yet, the $\bt$-dependence of the $\Mdisk$-$\jdisk$ relation in late-type systems (\S\ref{subsection_independent_models}), rules out the tempting idea that low-$j$ material simply migrates towards the bulge until the surrounding disk satisfies a certain bulge-independent criterion, such as a universal stability threshold. A more dynamic explanation is needed to account for the $\bt$-dependence of $\jdisk$.

To uncover the origin of the $M$-$j$-$\bt$ relation, let us note that this relation is approximately a monotonic relation between $\bt$ and $jM^{-1}$, similarly for baryons and stars, since $\jb\Mb^{-1}\approx\js\Ms^{-1}$ according to \S\ref{subsection_stars_vs_baryons}. Therefore, understanding the $M$-$j$-$\bt$ relation reduces to understanding the quantity $jM^{-1}$ and its effect on bulge formation. As for the first step, it is easily shown that $jM^{-1}$ is a measure of the surface density. In fact, using \eq{japprox1}, the surface density scale $\Sn\propto MR^{-2}$ can be rewritten as $\Sn\propto M j^{-1}R^{-1}\vflat$. Assuming a constant velocity $\vflat=\Vh$ and using $\Vh\propto H\Rh$ (from Equations (\ref{eq_Vh_isothermal}) and (\ref{eq_Rh_isothermal})), gives $\Sn\propto HMj^{-1}\Rh R^{-1}$, where $\Rh$ is the halo radius. If $R\propto\Rh$ (corresponding to constant $\lambda$ and $\fj$), then
\be\label{eq_Sigma_jM}
	\Sn \propto H M j^{-1}.
\ee
Thus, $jM^{-1}$ scales inversely with the surface density (or the `concentration') of the galaxy baryons. In this way, our finding that the bulge mass fraction $\bt$ scales inversely with $jM^{-1}$, confirms earlier evidence \citep{Prieto1989} for a relation between the morphology of spiral galaxies and their mean surface density.

Less obvious is the physics behind the connection between the surface density and $\bt$. Assuming that the bulge forms from instabilities in the gas-rich protogalaxies, the characteristic bulge growth rate \smash{$\dot M_{\rm bulge}/M$} and the final bulge mass fraction $\bt$ are expected to decrease monotonically with the stability of the protogalaxy. Locally, the stability of a flat disk against Jeans instabilities is quantified by the parameter $Q=\sigma\,\kappa\,(3G\Sigma)^{-1}$ \citep{Toomre1964}, where $\sigma$ is the local velocity dispersion,
$\kappa$ is the orbital frequency, and $\Sigma$ is the local surface density. By extension, the mean stability of the disk is then characterized by a global parameter $\avg{Q}\propto\sigman\kappan\,\Sn^{-1}$, where $\sigman$, $\kappan$, and $\Sn$ are normalization factors of the dispersion, orbital frequency, and surface density, respectively. For circular orbits, $\kappan\propto\vflat R^{-1}\propto H\Rh\,R^{-1}$. Assuming again that $R\propto\Rh$, yields $\kappan\propto H$ and
\be\label{eq_avgQ}
	\avg{Q}\propto H\sigman\Sn^{-1}.
\ee
Substituting $\Sn$ in \eq{avgQ} for \eq{Sigma_jM}, the explicit $H$-dependence disappears and
\be\label{eq_Q_from_jM}
	\avg{Q}\propto \sigman j M^{-1}.
\ee
This derivation shows that, up to variations in $\sigman$, a basic CDM-based galaxy model coupled with an instability-driven bulge can qualitatively account for the monotonic relation between $j M^{-1}$ and $\bt$.

The detailed processes governing the in situ formation of bulges as a function of the $Q$-parameter, including the physics of the velocity dispersion $\sigma$, remain subject to numerical modelling. Recent high-resolution hydrodynamic simulations with radiative feedback \citep{Elmegreen2008,Bournaud2014} suggest that the semi-stable gas-rich progenitors of modern spiral galaxies partially collapsed into giant star-forming clumps, which survived the strong radiative feedback over time-scales required to spiral to the galaxy center by dynamical friction. According to \citeauthor{Bournaud2014}, this clump-feeding of the bulge can approximately account for the bulge mass of typical spiral galaxies in the local universe and explain the observed structure and outflows of clumps in galaxies at redshift $z\approx2$ \citep{Genzel2011}. However, the question whether giant clumps survive long enough to migrate to the galaxy center remains debated as summarized by \cite{Glazebrook2013}: simulations still allow for both short \citep{Genel2012} and long lifetimes \citep{Ceverino2012}, depending on the model assumptions, and the observations remain non-conclusive \citep{Genzel2011,Wuyts2012,Guo2012}. Details on clumps aside, the success of high-resolution hydrodynamic simulations with radiative feedback in explaining the structure of spiral galaxies is encouraging and suggests that such simulations might hold the key to explaining the $M$-$j$-$\bt$ relation. However, to date, such simulations still represent a major computational challenge (Section \ref{section_introduction}).

\subsection{Intuitive Summary of the $M$-$j$-$\bt$ Scaling}

In essence, the $M$-$j$-$\bt$ scaling can be explained from similarity considerations summarizing \S\ref{subsection_interpretation_cdm} and \S\ref{subsection_idea_for_explanation}. Assuming self-similarity in 3D, the mass $\Mh$ of a halo is proportional to its characteristic volume \smash{$\Rh^3$}. Newtonian gravity then implies a circular velocity \smash{$\Vh\propto(\Mh/\Rh)^{1/2}\propto\Rh$}. Thus,
\be\label{eq_fundamental_scalingsx}
	\Vh\propto\Rh\propto\Mh^{1/3}.
\ee
Given these relations and a fixed $\lambda$, the specific angular momentum is \smash{$\jh=\Jh/\Mh\propto(\lambda\Rh\Mh\Vh)/\Mh\propto\Mh^{2/3}$}. This scaling extends to $M$ and $j$ in baryons/stars, up to variations in the ratios $M/\Mh$ and $j/\jh$. Thus,
\be\label{eq_fundamental_jM}
	j\propto M^{2/3}.
\ee
The scatter of this relation due numerically predicted variations in $\lambda$, $M/\Mh$, and $j/\jh$, approximately covers the shaded region in \fig{jm_diagram_CDM} for local spiral galaxies.

If the bulge grows from disk instabilities set by the 2D surface density $MR^{-2}$, then $\bt$ scales monotonically with $MR^{-2}\propto M\Rh^{-2}\propto M j^{-1}$ (use Equations~(\ref{eq_fundamental_scalingsx}) and (\ref{eq_fundamental_jM})). Hence, spiral galaxies of fixed $\bt$ satisfy
\be\label{eq_fundamental_jM2}
	j \propto M
\ee
with a proportionality factor that decreases monotonically with increasing $\bt$.

In brief, late-type galaxies scatter around a mean relation \smash{$j\propto M^{2/3}$}, representing 3D self-similarity (fixed volume density profile), while any subsample of fixed $\bt$ follows a relation $j\propto M$, representing 2D self-similarity (fixed surface density profile). Together, these scalings naturally explain why the bulge fraction of spiral galaxies tends to increase with their mass (c.f.~\fig{jm_diagram_CDM} gray shading versus solid lines).\\


\section{Conclusions}\label{section_conclusion}

This paper presented the first precision measurements (a few percent statistical uncertainty) of the specific angular momentum $j$ in stars and baryons (stars, atomic gas, and molecular gas) in nearby spiral galaxies. The study relies on all 16 spiral (Sab-Scd) galaxies of the THINGS sample with stellar and cold gas surface densities published by \cite{Leroy2008}. They cover baryon masses $\Mb$ of $10^9-10^{11}\msun$ and bulge mass fractions $\bt$ (=B/T) up to $0.32$, representative of most galaxies in the local universe \citep{Weinzirl2009}. The relations between $M$ (for baryonic $\Mb$ or stellar $\Ms$), $j$ (for $\jb$ or $\js$), and morphology were determined with unprecedented accuracy. The key findings are as follows.
\begin{itemize}
	\item $M$, $j$, and $\bt$ are strongly and irreducibly correlated. Their mean relation given in Equations (\ref{eq_fit_bt_from_M_and_j}) and (\ref{eq_fit_j_from_M_and_bt}) and visualized in \fig{MJB} is consistent with no intrinsic scatter.
	\item For a fixed $\bt$, the residual scaling is $j\propto M^\alpha$ with $\alpha\approx1$, thus $\bt$ varies monotonically with $j M^{-1}$. The exponent $\alpha\approx1$ is larger than those found by \cite{Romanowsky2012} for late-type galaxies of a fixed Hubble type. It is also larger than the exponent $\alpha\approx2/3$ obtained for all late-type galaxies, without fixing $\bt$. This explains why $\bt$ tends to increase with the mass of spiral galaxies (c.f.~\fig{jm_diagram_CDM} gray shading versus solid lines).
	\item The relations $\Mb$-$\jb$-$\bt$ and $\Ms$-$\js$-$\bt$ are very similar with fitting parameters consistent within their uncertainties. This similarity is partially coincidental and holds despite the fact that cold gas contributes significantly (30\%-40\%) to the baryon angular momentum $\Jb$ with a specific angular momentum about twice that of stars.
	\item The $M$-$j$-$\bt$ relation persists, when considering only the contribution to $M$ and $j$ from the disk without the bulge: the disk `knows' about the bulge via its angular momentum. Therefore, it is impossible to explain the $M$-$j$-$\bt$ relation of spiral galaxies from independent $M$-$j$ relations of the disk and bulge.
	\item The fundamental plane (FP) of spiral galaxies arises when the ($M,j$)-plane is mapped into 3D $(L,R,\vflat)$-space via Equations~(\ref{eq_fp_transformations}). Therefore, the FP and its projections, such as the Tully-Fisher relation, can be explained from the $M$-$j$ relation.
\end{itemize}

\cite{Koda2000} wrote ``We hypothesize that the 2D distribution [in the FP] implies the existence of two dominant physical factors in spiral galaxy formation ...''. This work suggests that mass and angular momentum are the two fundamental factors. With hindsight, the tight relation between $M$, $j$, and morphology, and similar relations for early-type galaxies \citep{Cappellari2011}, justifies the historical classification of galaxies by stellar mass and Hubble type. As IFS-based measurements of $j$ become easier, this historical classification might be substituted for a more fundamental and physically motivated classification by $M$ and $j$.\\


\section*{Acknowledgements}
D.O.~thanks Aaron Romanowsky, Camille Bonvin and Martin Bruderer for their helpful advice. We are grateful to the THINGS, SINGS, GALEX, HERACLES, and BIMA SONG teams for making their data publicly available. This research has made use of the NASA/IPAC Extragalactic Database (NED), which is operated by the Jet Propulsion Laboratory, California Institute of Technology, under contract with the National Aeronautics and Space Administration.


\appendix

\section{A.~Decomposition in disk and bulge}\label{appendix_disk_bulge}

\begin{figure*}[h]
 	\centering
	\includegraphics[width=8.6cm]{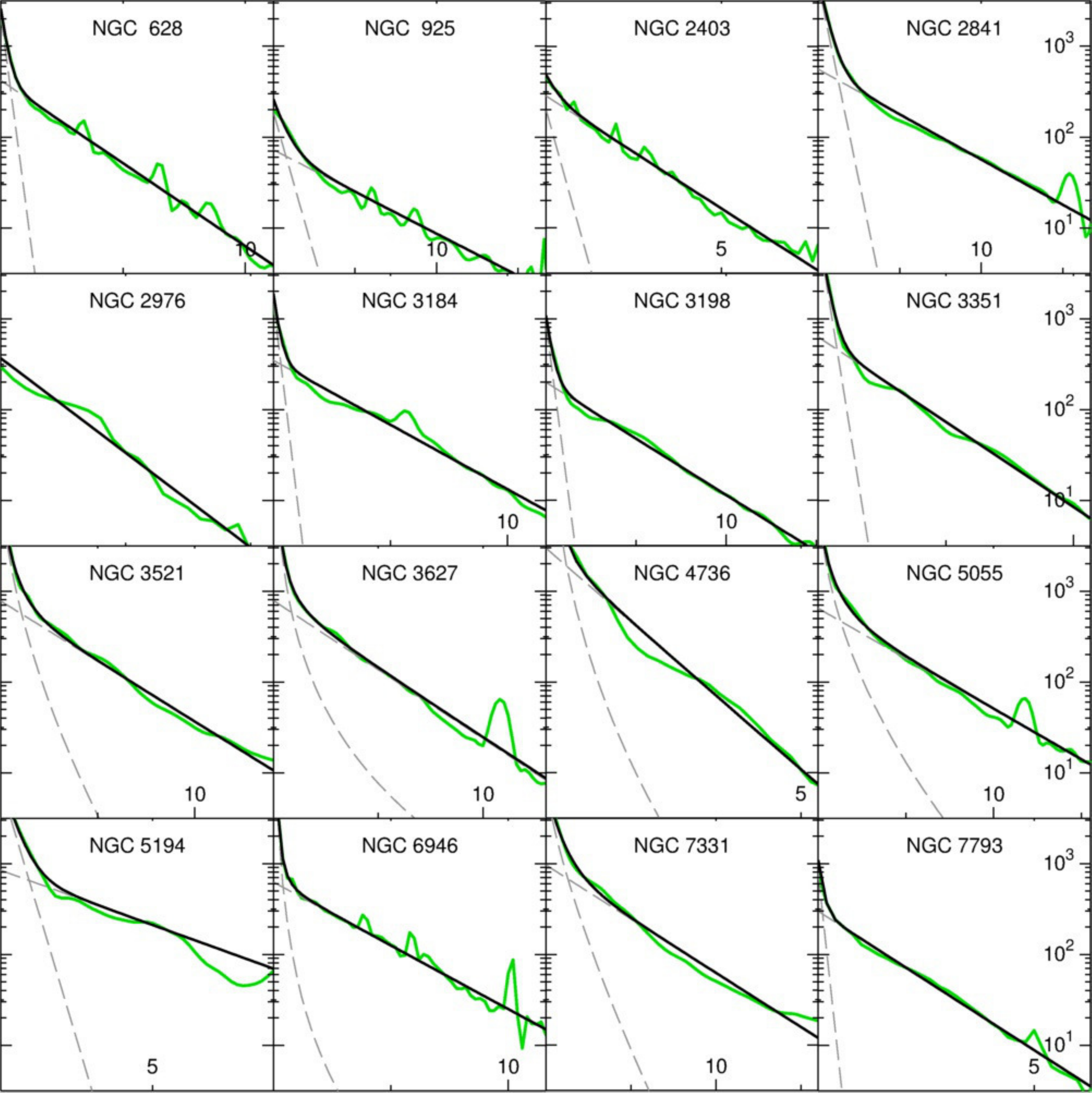}
	\caption{Decompositions of the stellar surface density profiles $\Ss(r)$ into disks and bulges. Green solid lines show the measured $\Ss(r)$, black solid lines show the fits $\Sigma_{\rm fit}(r)$ of \eq{disk_bulge}, and gray dashed lines represent the two components $\Sigma_{\rm d}(r)$ and $\Sigma_{\rm b}(r)$.\col}
	\label{fig_disk_bulge}
\end{figure*}

For each of the 16 galaxies, the stellar mass fraction $\bt$ of the `bulge' is calculated by fitting $\Ss(r)$ with a model composed of an exponential function for the disk (d) and a S{\'e}rsic profile \citep{Sersic1963} for the bulge (b),
\be\label{eq_disk_bulge}
	\Sigma_{\rm fit}(r) = \underbrace{k_{\rm d}\exp\bigg[-\frac{r}{R }\bigg]}_{\Sigma_{\rm d}(r)}+\underbrace{k_{\rm b}\exp\bigg[-\left(\frac{r}{r_{\rm b}}\right)^{1/n}\bigg]}_{\Sigma_{\rm b}(r)},
\ee
where $k_{\rm d}$, $R $, $k_{\rm b}$, $r_{\rm b}$, and $n>1$ (the `S{\'e}rsic index') are free parameters. Those are fitted to $\lg\Ss(r)$ using a robust fitting method \citep{Street1988} on the interval $r\in[0,\min(5\Rs,\Rmax)]$, where $\Rs$ is the disk scale radius determined by \cite{Leroy2008} and given in \tab{objects}, and $\Rmax$ is the maximal radius to which measurements for $\Ss(r)$ were published by \citeauthor{Leroy2008}. The fits $\Sigma_{\rm fit}(r)$ and their components $\Sigma_{\rm d}(r)$ and $\Sigma_{\rm b}(r)$ are plotted in \fig{disk_bulge}. Given those fits, the bulge mass fractions become $\bt=\int dr\,r\,\Sigma_{\rm b}(r)\,/\int dr\,r\,\Sigma_{\rm fit}(r)$. The standard errors of $\bt$ are typically around 0.02 as determined from multiple resampling of the data \citep{Efron1993}.


\section{B.~Angular momentum measurement}\label{appendix_J}

\begin{figure*}[t]
 	\centering
	\includegraphics[width=8.3cm]{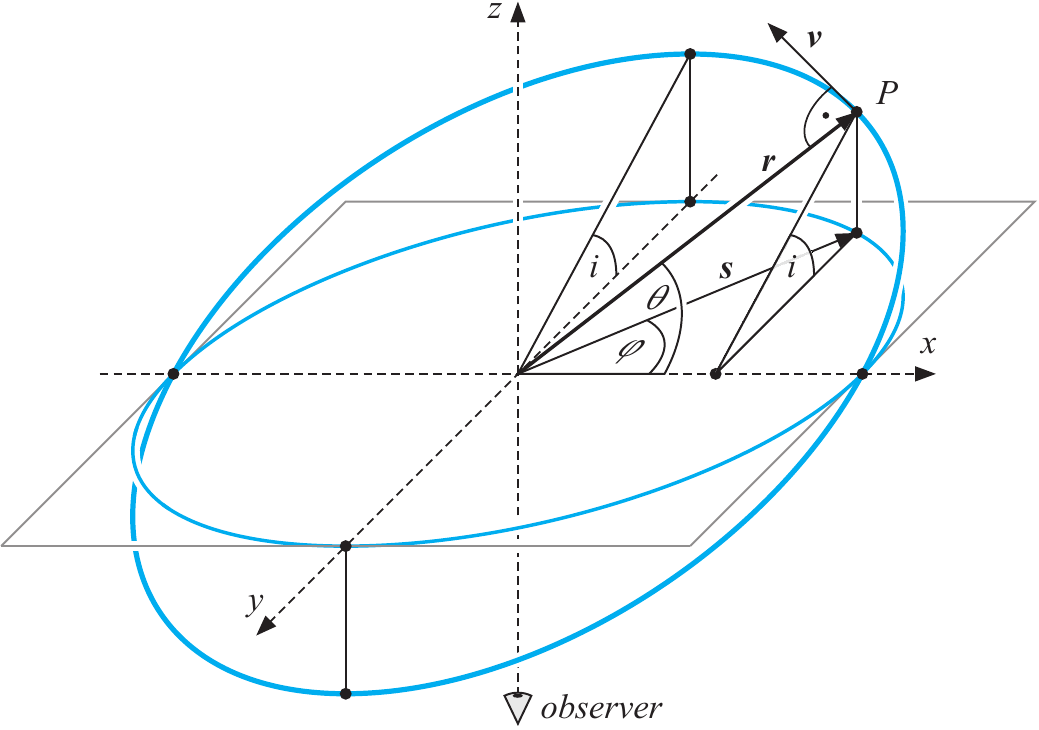}
	\caption{Schematic visualization of a particle $P$ on a circular orbit inclined against the line-of-sight.\col}
	\label{fig_deprojection}
\end{figure*}

Let us consider a flat galaxy with circular orbits, tilted against the observer by the inclination angle $i$, as shown in \fig{deprojection}. Here, this inclination $i$ is assumed to be known, since adopted from \citealp{Leroy2008}, but otherwise it can be determined from fits to the kinematic maps or from the minor-to-major axis ratio of the galaxy \citep[e.g.,][]{Obreschkow2013b}. Any orbiting point $P$ of mass $dM$ has a position vector $\mathbf{r}$ and a velocity vector $\mathbf{v}\perp\mathbf{r}$. The scalar angular momentum of the disk is given by
\be\label{eq_J}
	J=\left|\int dM~\mathbf{r}\times\mathbf{v}\,\right|=\int dM\,r\,v = \int_0^\infty dr~r^2\int_0^{2\pi}d\theta~\Sigma(r,\theta)~v(r,\theta),
\ee
where $\Sigma$ denotes the mass surface density of a specific baryonic component (e.g., stars). Upon assuming that the orbital velocity $v$ does not depend on $\theta$, or at least that variations of $v$ with $\theta$ are uncorrelated to the variations of $\Sigma$ with $\theta$ -- an assumption found correct at the 1\% level -- the second integral can be separated to $2\pi\Sigma(r)v(r)$ with $\Sigma(r)\equiv(2\pi)^{-1}\int_0^{2\pi}d\theta~\Sigma(r,\theta)$ and $v(r)$ being some azimuthally averaged mean of $v(r,\theta)$. Thus,
\be\label{eq_J_radial}
	J = 2\pi \int_0^\infty dr~r^2~\Sigma(r)~v(r).
\ee
Note that this simplification to radial profiles only does \emph{not} require or assume $\Sigma(r,\theta)$ to be invariant of $\theta$. To evaluate $J$ via \eq{J_radial}, $v(r)$ is needed, which requires measurements of $r$ and $v$ across the galaxy. However, these variables are not directly observable. Instead, for any pixel in the \ha\ maps (\fig{things}, left), one measures the projected radius $s$, its projected azimuth $\varphi$, i.e., the angle between the major axis and $\mathbf{s}$, and the recession velocity $v_z$. It is therefore necessary to calculate $r$ and $v$ from $s$, $\varphi$, $v_z$ and $i$. These relations are easily derived from \fig{deprojection} using basic trigonometry. Evoking the Pythagorean theorem,
\be\label{eq_deprojection_r}
	r = \big(r_x^2+r_y^2+r_z^2\big)^{1/2}=\big[(s\cos\varphi)^2+(s\sin\varphi)^2+(s\sin\varphi\tan i)^2\big]^{1/2}=s\big(\cos^2\varphi+\sin^2\varphi\cos^{-2}i\big)^{1/2}.
\ee
If $i=90^\circ$, the galaxy aligns with the $(x,z)$-plane and similarity implies $r_x/r=v_z/v$. As $i$ decreases, $r_x$, $r$ and $v$ remain unchanged, while $v_z$ must be substituted for $v_z\sin^{-1}i$, thus $r_x/r=v_z/(v\sin i)$. Using \eq{deprojection_r}, $v$ then solves to
\be\label{eq_deprojection_v}
	v = \frac{r}{r_x}\,\frac{v_z}{\sin i} = \frac{\big(\cos^2\varphi+\sin^2\varphi\cos^{-2}i\big)^{1/2}}{\cos\varphi\sin i}\,v_z \equiv C(\varphi,i)\,v_z,
\ee
where we introduced the local velocity deprojection factor $C(\varphi,i)$. Note that the term $r/r_x$ cannot be simplified to $(1+\tan^2\varphi\cos^{-2}i\big)^{1/2}$, since $(\cos^2\varphi)^{1/2}\neq\cos\varphi$ if $\cos\varphi<0$.

\begin{figure*}[t]
	\centering
	\begin{tabular}{cc}
		\begin{overpic}[width=8.6cm]{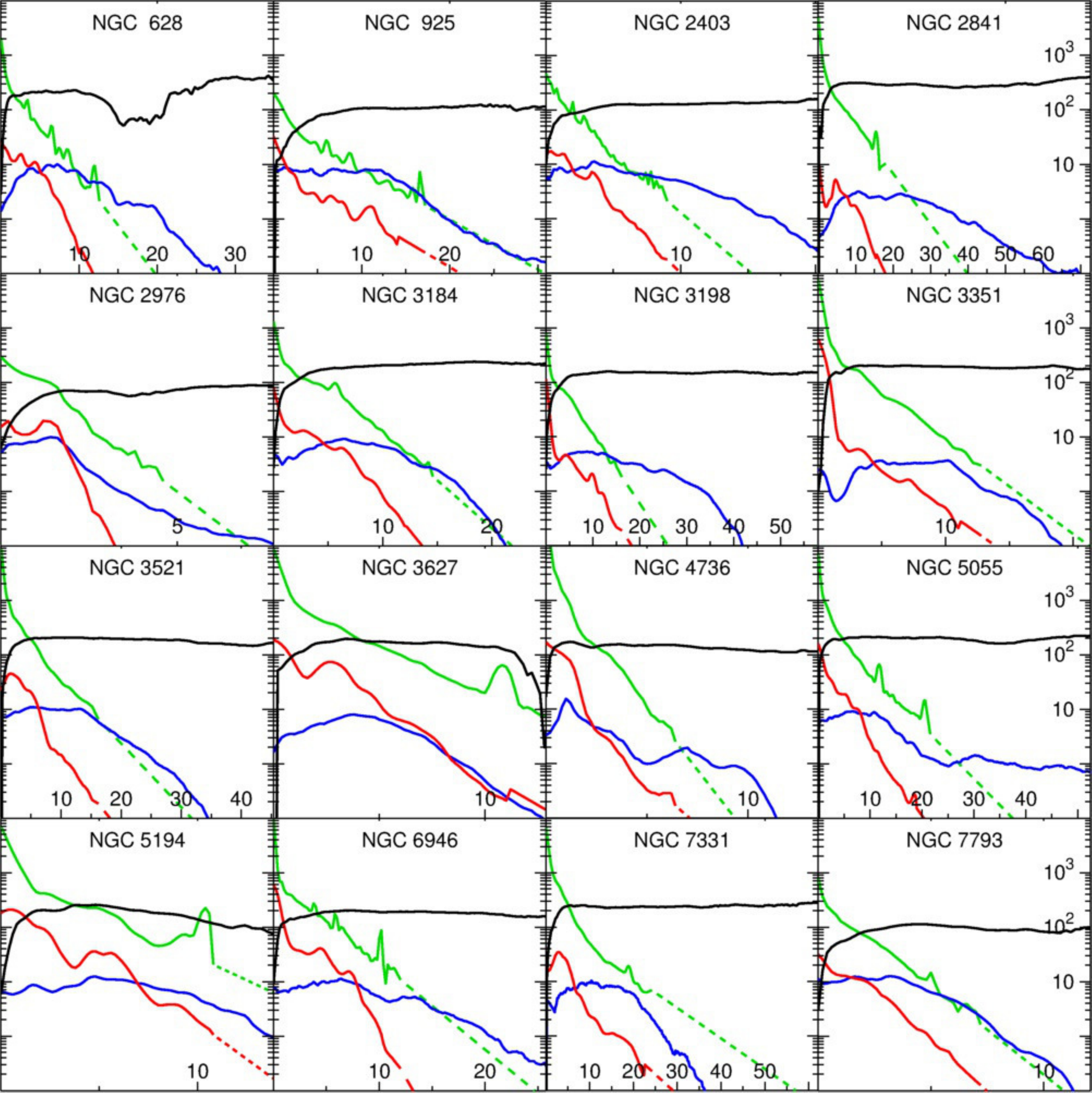}\end{overpic} & 
		\begin{overpic}[width=8.6cm]{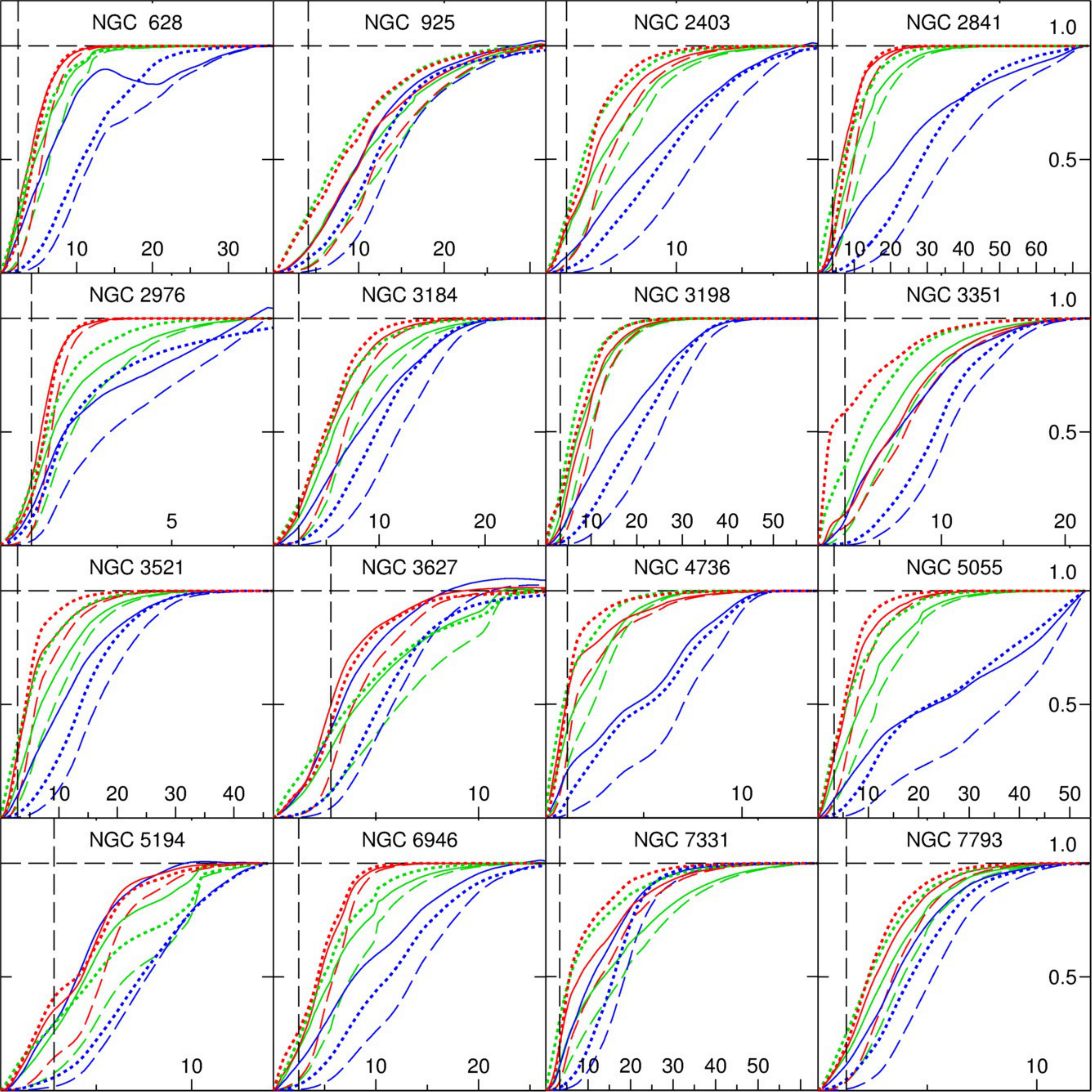}\end{overpic}
	\end{tabular}
	\caption{Left: inclination-corrected, azimuthally averaged circular velocity profiles $v(r)$ in units of $\kms$ (black), extracted from the \ha\ velocity maps; and inclination-corrected, azimuthally averaged mass surface densities of stars (green), \ha\ (blue), and \hm\ (red) in units of $\msun\rm pc^{-2}$. Solid colored lines represent the measurements adopted from \cite{Leroy2008} for stars and \hm\ and extracted from the \ha\ intensity maps of \cite{Walter2008} for \ha. Dashed lines represent exponential extrapolations, where no data is available. Right: normalized cumulative functions of mass $M(r)$ (dotted, \eq{cumM}), angular momentum $J(r)$ (dashed, \eq{cumJ}), and specific angular momentum $j(r)$ (solid, \eq{cumj}). Different colors represent stars (green), \ha\ (blue), and \hm\ (red).\col}
	\label{fig_profiles}
\end{figure*}

To evaluate the function $v(r)$ of a real galaxy, Equations (\ref{eq_deprojection_r}) and (\ref{eq_deprojection_v}) are applied to every pixel $k$ in the 2D \ha\ map ($2048\times2048$ pixels for NGC 2403, $1024\times1024$ pixels for the other 15 galaxies). Using both the intensity (moment 0) and velocity (moment 1) maps, each pixel $k$ is given a value $\{I_k,r_k,v_k\}$, where $I_k$ denotes the intensity. The data is then binned into different radii, equally spaced by 100~pc. In every bin, the mean velocity is calculated as the mean of the pixel velocities, weighted by intensity and the variance $C(\varphi,i)^{-2}$ of the deprojection error,
\be
	v_{\rm bin} = \frac{\sum_{k\in\rm bin}I_k C(\varphi,i)^{-2} v_k}{\sum_{k\in\rm bin}I_k C(\varphi,i)^{-2}}.
\ee
This results in a discrete function $v(r)$ known at steps of 100~pc. In turn, the different density profiles $\Sigma(r)$ are given at 200~pc to 700~pc spacings. These profiles are re-gridded to 100~pc spacings using a spline-interpolation in order to multiply them with $v(r)$ in the computation of $J$.

\fig{profiles} (left) shows the radial surface densities $\Sigma(r)$ of stars, \ha, and \hm~(including helium) together with the extracted velocity profiles $v(r)$. The corresponding normalized cumulative functions of mass and angular momentum are shown in \fig{profiles} (right). They are defined as
\begin{eqnarray}
	M(r) & = & 2\pi \int_0^r dr'~r'~\Sigma(r'),\label{eq_cumM} \\
	J(r) & = & 2\pi \int_0^r dr'~r'^2~\Sigma(r')~v(r'),\label{eq_cumJ} \\
	j(r) & = & J(r)/M(r).\label{eq_cumj}
\end{eqnarray}
Models of $M(r)$, $J(r)$, and $j(r)$ based on an exponential disk (see \eq{jR} for $j(r)$) are used to estimate the uncertainty of $M$, $J$, and $j$ due to the finite size of the maximal observable radius $\Rmax$. In the limit of this exponential model, the relative difference between $j(\Rmax)$ and $j$ is $45.6\%$, $8.7\%$, and $0.2\%$, if $\Rmax/R=2$, 5, and 10, respectively, where $R$ is the exponential scale radius of $\Sigma(r)$. Since most galaxies studied here were measured to $\Rmax\approx10R$ for stars and \hm\ (with extrapolations to the \ha~radii $R_{\rm max,HI}\approx14R$), the values $\js$ and $\jhm$ are converged to less than 1\%. Explicit fits of \eq{jR} to the measured $j(r)$ suggest that $\jb$, $\js$, $\jhm$ are converged at the 1\% level, while $\jg$, $\jha$ are converged at the 10\% level. Only in the case of NGC 5055 $\jha$ might be 30\% larger than measured, but even in this case the baryonic $\jb$ changes by less than 10\%. Additional statistical and systematic uncertainties are discussed in \S\ref{subsection_j_measurement}. As a sanity check of the deprojection method, the Pearson correlation coefficient $c$ between the inclinations $i$ and the values $\jb$ was computed and revealed no significant correlation ($c\approx0.2$).


\section{C.~Multivariate linear regressions}\label{appendix_regressions}

The bivariate linear regression is a method to fit the linear equation
\be\label{eq_linear_relation_2d}
	y = k_1x+k_2
\ee
with free parameters $k_1$ and $k_2$ to a set of 2D data points. This regression is \textit{optimal} in the sense that it provides the \textit{most likely} linear relation for data that intrinsically lies on a linear relation, but has been scattered by uncorrelated Gaussian noise of known variance. This noise can apply to both dimensions and may be different for each data point. The bivariate linear regression is obtained by minimizing
\be\label{eq_bivariate_regression}
	\chi^2 = \sum_i\frac{(k_1x_i+k_2-y_i)^2}{k_1^2\sigma_{x,i}^2+\sigma_{y,i}^2},
\ee
where $(x_i,y_i)$ are the measured values and $\sigma^2_{x,i}$ and $\sigma^2_{y,i}$ are their variances in both dimensions.

In the same sense, the trivariate linear regression is the optimal method to fit the linear equation
\be\label{eq_linear_relation_3d}
	z = k_1x+k_2y+k_3
\ee
with free parameters $k_1$, $k_2$, and $k_3$ to a set of 3D data points. This regression is obtained by minimizing
\be\label{eq_trivariate_regression}
	\chi^2 = \sum_i\frac{(k_1x_i+k_2y_i+k_3-z_i)^2}{k_1^2\sigma_{x,i}^2+k_2^2\sigma_{y,i}^2+\sigma_{z,i}^2},
\ee
where $(x_i,y_i,z_i)$ are the measured values and $\sigma^2_{x,i}$, $\sigma^2_{y,i}$, and $\sigma^2_{z,i}$ their variances.

In this work, the $\chi^2$-minimization is performed using MATLAB's `fminsearch' function, which relies on the Nelder-Mead simplex algorithm as described by \cite{Lagarias1998}.\vfill


\end{document}